\documentclass[]{emulateapj}
\usepackage{apjfonts}
\newcommand{\be}{\begin{equation}}
\newcommand{\ee}{\end{equation}}
\newcommand{\bea}{\begin{eqnarray}}
\newcommand{\eea}{\end{eqnarray}}

\newcommand{\vol}{{\mathrm{Mpc}}^{-3}}

\newcommand{\Msun}{M_{\odot}}

\newcommand{\Mvir}{M_{\mathrm{vir}}}

\def\Mpc{\ {\rm Mpc}}

\newcommand{\ac}{\ifmmode{a_{\rm c}}\else$a_{\rm c}$\fi}
\newcommand{\zobs}{\ifmmode{z_{\rm o}}\else$z_{\rm o}$\fi}
\newcommand{\aobs}{\ifmmode{a_{\rm o}}\else$a_{\rm o}$\fi}
\newcommand{\Mobs}{\ifmmode{M_{\rm o}}\else$M_{\rm o}$\fi}

\slugcomment{Submitted to ApJ, 22 May 2008}
\shortauthors{CONROY AND WECHSLER}
\shorttitle{CONNECTING GALAXIES, HALOS, AND STAR FORMATION RATES}
\begin{document}

\title{ Connecting Galaxies, Halos, and Star Formation Rates across
  Cosmic Time }
\author{Charlie Conroy$^{1}$ \& Risa H. Wechsler$^{2}$}
\affil{$^{1}$Department of Astrophysical Sciences, Princeton
  University, Princeton, NJ 08544 \\ $^{2}$Kavli Institute for
  Particle Astrophysics \& Cosmology, Physics Department, and Stanford
  Linear Accelerator Center, Stanford University, Stanford, CA 94305}

\slugcomment{Submitted to ApJ, 22 May 2008}

\begin{abstract}

  A simple, observationally-motivated model is presented for
  understanding how halo masses, galaxy stellar masses, and star
  formation rates are related, and how these relations evolve with
  time.  The relation between halo mass and galaxy stellar mass is
  determined by matching the observed spatial abundance of galaxies to
  the expected spatial abundance of halos at multiple epochs ---
  i.e. more massive galaxies are assigned to more massive halos at
  each epoch.  This ``abundance matching'' technique has been shown
  previously to reproduce the observed luminosity- and
  scale-dependence of galaxy clustering over a range of epochs.  Halos
  at different epochs are connected by halo mass accretion histories
  estimated from $N$-body simulations.  The halo--galaxy connection at
  fixed epochs in conjunction with the connection between halos across
  time provides a {\it connection between observed galaxies across
    time}.  With approximations for the impact of merging and
  accretion on the growth of galaxies, one can then directly infer the
  star formation histories of galaxies as a function of stellar and
  {\it halo} mass.  This model is tuned to match both the observed
  evolution of the stellar mass function and the normalization of the
  observed star formation rate -- stellar mass relation to $z\sim1$.
  The data demands, for example, that the star formation rate density
  is dominated by galaxies with $M_{\rm star}\approx
  10^{10.0-10.5}\Msun$ from $0<z<1$, and that such galaxies over these
  epochs reside in halos with $M_{\rm vir}\approx10^{11.5-12.5}\Msun$.
  The star formation rate -- halo mass relation is approximately
  Gaussian over the range $0<z<1$ with a mildly evolving mean and
  normalization.  This model is then used to shed light on a number of
  issues, including 1) a clarification of ``downsizing'', 2) the lack
  of a sharp characteristic halo mass at which star formation is
  truncated, and 3) the dominance of star formation over merging to
  the stellar build-up of galaxies with $M_{\rm star}\lesssim
  10^{11}\Msun$ at $z<1$.

\end{abstract}


\section{Introduction}
\label{section:intro}

A fundamental goal of galaxy formation studies is to understand what
processes govern the stellar content and star formation histories of
galaxies.  A key piece of this puzzle is relating the stellar masses
and star formation rates of galaxies to the masses and formation
histories of their associated dark matter halos.  Ideally, one would
like to make this connection by understanding the physical mechanisms
responsible for it from first principles.  However, even the best
current physically-motivated models of galaxy formation rely on
significant approximations of unresolved physics.  These approaches,
based either on semi-analytic modeling \citep[e.g.][]{White91,
  Somerville99, Cole00, Hatton03, Springel01, Croton06, Bower06}, or
on hydrodynamical simulations \citep[e.g.][]{Cen92, Katz96,
  Springel03, Keres05} still have trouble reproducing many basic
observational results and suffer from serious uncertainties in the
physical ingredients of the models.  Although substantial progress has
been made in these modeling efforts in recent years, star formation
histories in these models and simulations are still sensitive to the
interactions between a number of relatively unconstrained physical
processes.

Recent observations have begun to measure the galaxy stellar mass
function \citep{Fontana04, Drory04, Bundy05, Borch06, Fontana06,
  Cimatti06, Andreon06} and the star formation rate \citep{Noeske07a,
  ZhengBell07} at high redshift, which complements more precise
measurements locally \citep[e.g.][]{Cole01, Bell03, Brinchmann04,
  Panter07, Salim07, Schiminovich07}.  At the same time, the evolution
of dark matter halos, including their abundance
\citep[e.g.][]{Warren06, Reed07}, substructures \citep{Kravtsov04,
  Gao04b, Reed05}, and merger and accretion histories
\citep[e.g.][]{Wechsler02}, are becoming ever better understood in the
context of the $\Lambda$CDM paradigm using numerical simulations.

Several methods have recently been developed that take advantage of
these advances to connect the observed galaxy population with dark
matter halos using more empirical methods.  The most popular of these,
known as halo occupation models, typically constrain the statistics of
how galaxies populate their host halos using galaxy clustering
statistics and space densities \citep[e.g.][]{Scoccimarro01,
  Berlind02, Bullock02, Zehavi04}.  An emerging alternative is to connect
galaxies to the underlying dark matter structure directly, under the
assumption that the stellar masses or luminosities of the galaxies are
tightly connected to the masses or circular velocities of dark matter
halos.  Throughout, this latter approach will be referred to as
halo ``abundance matching'' because galaxies of a given stellar
mass are matched to halos ({\em including subhalos}, which are halos that
orbit within larger halos) of the same number density or abundance.
This approach matches the observed stellar mass function by
construction, but has no other observational inputs.  Such an approach
provides an excellent match to a number of galaxy clustering
statistics at multiple epochs \citep{Kravtsov04, Tasitsiomi04, Vale04,
  Conroy06a, Berrier06, Vale06, Marin08}.

The idea of abundance matching galaxies with dark matter halos is not
new, and it has been applied to associate a variety of objects with
halos since the development of the CDM paradigm \citep[e.g.][]{Mo96c,
  Mo96b, Steidel98, Wechsler98}.  However, its successful
implementation as a predictive tool requires a full accounting of the
halo population, including the substructures that host galaxies, as
well as a full accounting of the evolution of the abundance of
galaxies as a function of their properties.  These elements have only
been in place quite recently.

Halo occupation models as well as abundance matching models have been
used primarily to understand the connection between galaxies and halos
at a fixed epoch, but recent work has begun to use these models to
investigate the evolutionary history of galaxies, by combining
information about the galaxy--halo connection at given epochs with
theoretical input on the evolution of dark matter halos
\citep{MWhite07, Conroy07b, Zheng07, Conroy08a}.  In this paper we
take the basic idea of abundance matching further, and use it to
understand the evolution of the stellar content of galaxies.  We use a
simple, analytic representation of this framework, which connects dark
matter halos to galaxies by matching their abundances, to understand
the build-up of stellar mass and the implied star formation rate of
galaxies as a function of mass.  We focus primarily on redshifts less
than one, where the observational results are most reliable, but we
expect the approach can be applied more widely and to earlier epochs
as observational results improve.

A complementary approach has recently been presented by
\cite{Drory08}.  While we use the measured galaxy stellar mass
function to connect galaxies to dark matter halos and infer the stellar mass
buildup and star formation rates of galaxies, they used the measured
star formation rates as a function of stellar mass, along with the
time derivative of the galaxy stellar mass function, to infer the
galaxy merger rate. 

The elements of our model are described in detail in
$\S$\ref{sec:themodel}; $\S$\ref{sec:imp} presents our primary
results, including comparisons to observations.  We discuss some of
the implications of our model in $\S$\ref{sec:imp} and summarize in
$\S$\ref{sec:disc}.  Throughout a flat, $\Lambda$CDM cosmology is
assumed with the following parameters:
$(\Omega_m,\Omega_\Lambda,\sigma_8) = (0.24,0.76,0.76)$, and $h=0.7$
where $h$ is the Hubble parameter in units of $100$ km s$^{-1}$
Mpc$^{-1}$.  These cosmological parameters are consistent with the 3rd
year {\it WMAP} estimates \citep{Spergel07}.  A \citet{Chabrier03}
initial mass function (IMF) is adopted throughout.

\section{The Model}
\label{sec:themodel}

This section describes the details of our model.  We start with a
brief overview, and then move to a discussion of the halo mass
function and galaxy stellar mass functions in $\S$\ref{sec:hmf} and
\ref{sec:gsmf}.  The method used to assign galaxies to halos is
outlined in $\S$\ref{sec:am}, followed by a description of the
approach used to connect galaxies and halos across epochs in
$\S$\ref{sec:mah}.  Introducing a simple estimate for the effect of
galaxy mergers and accretion in $\S$\ref{sec:merge} then allows us to
compute star-formation histories of galaxies, as discussed in
$\S$\ref{sec:sfr}.

\subsection{Overview}
\label{sec:over}

The model described in detail in the following sections is an
extension of previous modeling efforts that have been shown to
successfully reproduce an array of data from $z\sim5$ to the
present\citep{Kravtsov04, Tasitsiomi04, Vale04, Conroy06a, Berrier06,
  Vale06, Marin08}.  The first step in our approach is to match the
observed abundances of galaxies as a function of stellar mass with the
expected abundance of dark matter halos. This step effectively assigns
the most massive galaxies to the most massive halos monotonically and
with no scatter.  Since we include dark matter subhalos, which are
halos orbiting within larger halos, we automatically include galaxies
that would be observationally classified as satellites, although they
are sub-dominant by number ($\sim10-30$\% of the galaxies are
satellites at any epoch).  Thanks to parameterizations of both the
evolution of the observed galaxy stellar mass function and of the
theoretical halo mass function, this connection between galaxies and
dark matter halos can be determined continuously from $z\sim2$ to
$z\sim0$.

The novel feature of our approach, compared to previous work, is the
use of average dark matter mass accretion histories to connect the
relations between halos and galaxies across time.  $N$-body
simulations suggest that the average dark matter halo growth is a
simple function of its mass \citep{Wechsler02}; thus, a halo at any
given epoch can be connected to its typical descendants at later
epochs.  With the connection between galaxies and halos determined at
each epoch, the connection between halos across time implies an
\emph{average connection between galaxies across time}.  At this stage
the model produces the average stellar mass growth of galaxies as a
function of both galaxy and halo mass.  Since we use
observationally-derived galaxy stellar mass functions as input, the
connection is effectively one between observed galaxies at different
epochs.

The final step is to differentiate these average stellar mass growth
curves to infer the average mass-growth rates of galaxies.  The
complication here is separating the growth due to star formation from
that due to merging/accretion of other stellar systems.  We introduce
simple estimates of the contribution due to merging that should
bracket the possible effects of merging.  This model then allows us to
determine the average star formation rates of galaxies as a function
of their {\em halo mass} and redshift, which provides a key constraint
on galaxy formation models.  The following sections describe this
framework in further detail.

\subsection{The halo mass function}
\label{sec:hmf}

We use the cosmology- and redshift-dependent halo mass function given
by \citet{Warren06} and transform their masses to $\Mvir$ using an NFW
\citep{NFW97} density profile with the concentration--mass relation
from \cite{Bullock01}, assuming the updated model parameters given by
\cite{Wechsler06}.  Our definition of the virial radius corresponds to
region with density contrast $\Delta_{\rm{vir}} = 18\pi^2+82x-39x^2$
with respect to the mean matter density, where $x\equiv \Omega(z)-1$
\citep{Bryan98}.  At $z=0$, $\Delta_{\rm{vir}} = 337$, and at high
redshift $\Delta_{\rm{vir}}$ asymptotes to $180$.

The halo mass function provided by \citet{Warren06} only considers
distinct halos, not the substructure within these distinct halos.
Substructure as defined herein consists of halos whose centers are
within the virial radii of larger halos, denoted subhalos.  Distinct halos, in contrast, are those halos whose
centers are not within any larger halos.

We assume that the subhalo fraction is described by
\noindent
\be
f_{\rm sub}\equiv \frac{n_{\rm sub}}{n_{\rm tot}} = 0.2
-\frac{0.1}{3}\,z,
\ee
\noindent
independent of distinct halo mass, which provides a reasonable fit to
data from simulations (see e.g., Fig. 1 of \citealt{Conroy06a} we
don't include the moderate decrease of $f_{\rm sub}$ with increasing
mass indicated by simulation data, but this would have a small effect
on our results).  Note that the subhalo fraction is defined with
respect to the mass of the subhalos at the epoch of their accretion.
This mass, rather than the present subhalo mass, has been shown to
better correlate with observed galaxy properties \citep{Conroy06a,
  Berrier06}.  We thus derive an approximate halo mass function that
includes both distinct halos and subhalos using this fraction.  The
results presented below are fairly insensitive to this fraction
because it is small; we include it for completeness.  Throughout, we
refer to both distinct halos and subhalos as halos.

\begin{figure}[t]
\plotone{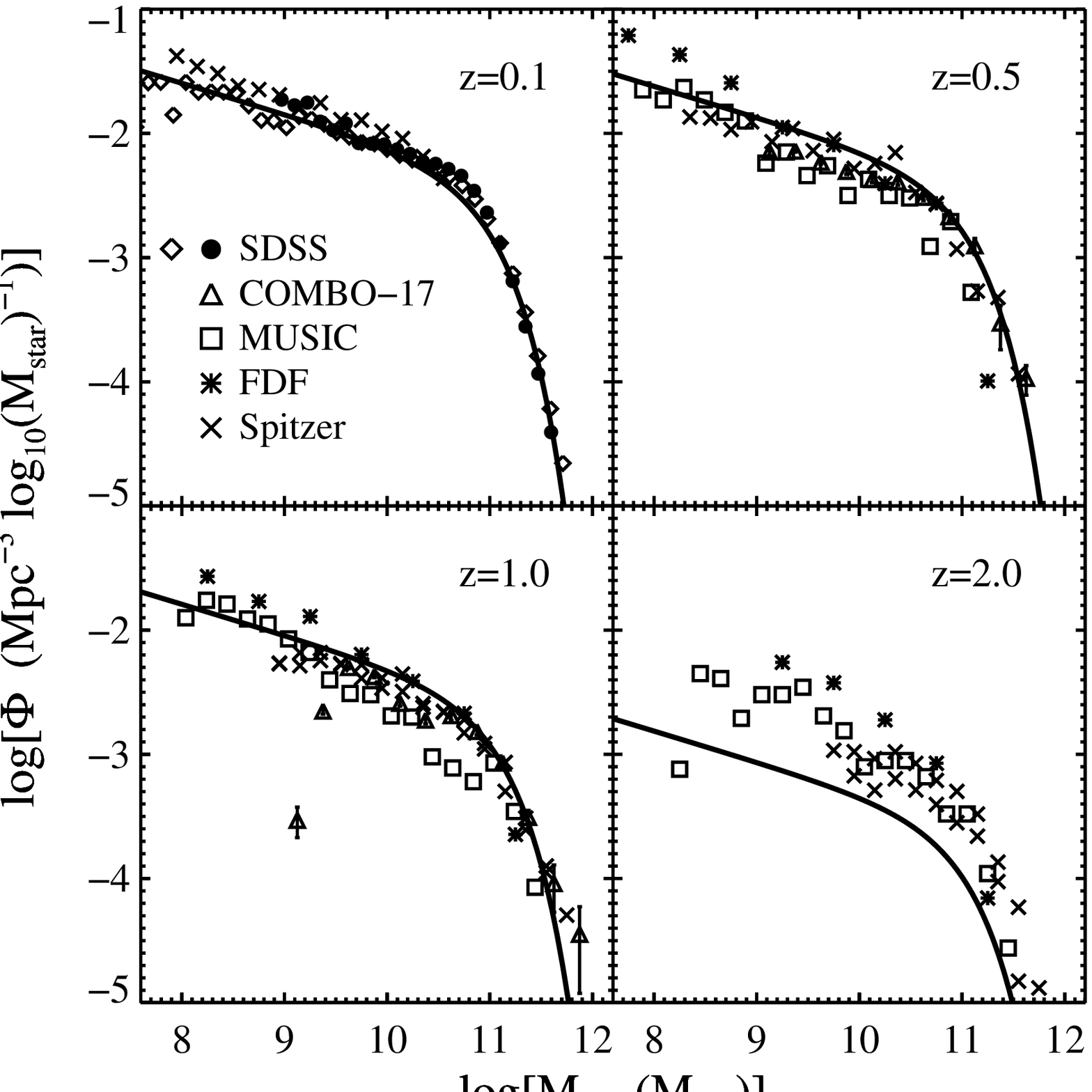}
\vspace{0.5cm}
\caption{Evolution of the galaxy stellar mass function from $z\sim2$
  to $z\sim0$.  Our fiducial model for the evolution of the mass
  function (\emph{lines}) is compared to the following observational
  results from the literature: \citet[][SDSS; {\it circles}]{Bell03},
  \citet[][SDSS; {\it diamonds}]{Panter07} \citet[][FDF]{Drory05},
  \citet[][COMBO-17]{Borch06}, \citet[][{\it
    Spitzer}]{Perez-Gonzalez08}, and \citet[][MUSIC]{Fontana06}.  The
  disagreement between model and data at $z=2$ is discussed in
  $\S$\ref{sec:highz}.}
\label{fig:gsmf}
\vspace{0.5cm}
\end{figure}

\subsection{The galaxy stellar mass function}
\label{sec:gsmf}

At each redshift, the number density $\phi(M,z)$ of galaxies with
stellar mass $M_*$ is assumed to be described by a Schechter function,
\be \phi(M,z) = \phi^*(z) \bigg(\frac{M}{M^*(z)}\bigg)^{\alpha^*}
\exp\bigg(-\frac{M}{M^*(z)}\bigg), \ee
\noindent
where the free parameters $\phi^*(z)$, $\alpha^*(z)$ and $M^*(z)$ are,
in principle, functions of redshift.  We take the evolution of $M^*(z)$ 
to be:
\noindent
\be
{\rm log}[M^*(z)/\Msun] = 10.95 + 0.17 \, z -0.07 \,\, z^2,
\label{sche_mstar}
\ee
\noindent
which is similar to the form advocated by \citet{Fontana06}.  Note
that the evolution in $M^*$ implied from the above formula is mild at
$z<2$.  Since the constraints on $\alpha^*$ are weak at higher
redshift, we assume for simplicity that it does not evolve:
\noindent
\be
\alpha^* = -1.25,
\label{sche_alpha}
\ee
\noindent
which is consistent within the errors with available data to $z\sim2$
\citep{Fontana06}.

The evolution of $\phi^*(z)$ raises a subtle but important issue.
Various authors have measured $\phi^*$ in redshift bins to
$z\sim2$ and then proceeded to fit the observed $\phi^*(z)$ to a
function of the form $\phi^*(z)\propto (1+z)^{-\beta}$.  However, it
is clear that, modulo small evolution in $M^*$ and $\alpha^*$ (and
corrections due to stellar mass loss; cf. $\S$\ref{sec:sfr}), the
time-derivative of $\phi^*(z)$ is simply the cosmic star formation
rate (SFR) density.  The functional form above, for typical adopted
values of $\beta=1-3$, results in an {\it increasing} SFR density at
late times.  This is not observed \citep[e.g.][]{Hopkins04}.

In order to alleviate this tension, we have chosen to constrain the
evolution in $\phi^*(z)$ by requiring both that it reproduce the
observed evolution in the stellar mass function and that its
derivative match the normalization of the observed star formation rate
-- stellar mass relation to $z\sim1$.  After some experimentation with
different functional forms, we adopt the following evolution of
$\phi^*$:
\noindent
\be
\phi^*(z)=  2\times 10^{-3}\, e^{-0.5z^{2.5}} \,\,\vol.
\label{sche_phi}
\ee
\noindent
Note that our parameterization is by no means unique.  We have simply
attempted to match the observed cosmic star formation rate density
implied by our model and the observed stellar mass functions by
adjusting the form of $\phi^*(z)$.  This functional form is similar to
that given in \citet{Wilkins08} who proposed $\phi^*(z) = 2.5\times
10^{-3}\, e^{-0.68z^{1.2}} \,\,\vol$ as the best-fit to a variety of
stellar mass function data.

The zero-points of the Schechter parameters approximately reproduce
the local set of parameters determined by a variety of authors
\citep{Cole01, Bell03, Wang06, Panter07}.  Figure \ref{fig:gsmf}
compares the evolution of the galaxy stellar mass function in our
model to various observational estimates. Our adopted Schechter
parameters somewhat overpredict the abundance of low-mass galaxies at
$z=0.5$ and underpredict the abundances of all galaxies at $z=2$.  The
latter disagreement is discussed in $\S$\ref{sec:highz}.

There is a second, perhaps more serious, tension created by comparison
of the evolution of the stellar mass function and the global SFR
density.  At $z\gtrsim1$ the integral of the star formation rate
density does not equal the stellar mass density \citep{Nagamine06,
  Hopkins06b, Perez-Gonzalez08, Wilkins08}.  This tension can largely
be removed if the IMF evolves with redshift \citep{Dave08, Wilkins08}
because the SFR probes the high-mass end of the IMF while the bulk of
the stellar mass is contained in low-mass stars.  An evolving IMF at
$z\gtrsim1$ is also suggested by recent work on the evolution of the
mass-to-light ratio of elliptical galaxies \citep{vanDokkum08} and the
abundance patterns of metal-poor stars \citep{Lucatello05,
  Tumlinson07a, Tumlinson07b}.  Whether this is the ultimate solution,
or whether the solution lies in a more mundane systematic error in one
of the measured quantities is not currently clear.  Because of this
tension at high redshift, we focus our analysis below $z\sim1$, where
the cosmic SFR density and stellar mass density are consistent with
each other assuming a non-evolving IMF.

\subsection{Abundance matching: from halos to galaxies}
\label{sec:am}

We assume that every galaxy resides in a dark matter halo and that
there is a tight connection between the stellar mass of a galaxy and
the mass of its associated dark matter halo.  In the limit of zero
scatter between galaxy and halo mass, halos of a given mass can be
connected to galaxies of a given stellar mass by matching their
abundances directly:
\noindent
\be
\label{eqn:v2l}
n_g(>{M_{{\rm star},i}}) = n_h(>{M_{{\rm vir},i}}),
\ee
\noindent
where $n_g$ and $n_h$ are the galaxy and halo mass functions,
respectively (note that the halo mass function here includes both
distinct halos and subhalos).  In effect, this prescription assigns
the most massive galaxies to the most massive halos monotonically.
Although the assumption of zero scatter is idealized, several recent
works indicate that this scatter is small, with $\sim 0.15$ dex of
scatter in galaxy luminosity at fixed mass \citep{Zheng07, vdB07,
  Hansen07, Wechsler08}.  As shown in \citet{Tasitsiomi04}, scatter
only effects the halo-stellar mass relation at the high mass end, and
is in the sense that, at fixed galaxy mass, the mean halo mass
decreases with increasing scatter.

Since more massive halos are more strongly clustered at all epochs,
this mapping implies that more massive/more luminous galaxies will
also be more strongly clustered than less massive/less luminous ones,
in qualitative agreement with a variety of clustering measurements
\citep[e.g.][]{Zehavi05, Coil06b, Li06, Meneux08}.  This simple
approach is surprisingly successful at quantitatively matching an
array of observations at multiple epochs and scales including
mass-to-light ratios, clustering measurements, and close pair counts
\citep{Kravtsov04, Tasitsiomi04, Vale04, Vale06, Berrier06, Conroy06a,
  Marin08}, confirming that it can be used with confidence herein.

\subsection{Mass accretion histories: from halo growth to galaxy growth}
\label{sec:mah}

Analysis of cosmological $N$-body simulations has shown that halo mass
growth can be described by a simple functional form
\citep{Wechsler02}:
\noindent
\be 
M_{\rm vir}(a) = \Mobs {\rm exp} \left[-2\ac \left(\frac{\aobs}{a}-1\right)\right],
\label{eq:fit2}
\ee
\noindent
where $a=(1+z)^{-1}$, $\Mobs$ is the mass of the halo at the redshift
of observation $a_{\rm o}$, and $a_c$ is the average formation scale
factor of the halo, the single free parameter in the functional form
defined above.  Following \citet{Wechsler02}, we adopt the following
parameterization of $a_c$, which provides a good fit to $N$-body
simulations:
\noindent
\be
a_c(M_{\rm vir}) = \frac{4.1}{c(M_{\rm vir})(1+z)}.
\ee
\noindent
where $c(M_{\rm vir})$ is the halo concentration$-$ halo mass relation
at $z=0$.  We use the model given by \cite{Bullock01} and the updated
parameters provided in \cite{Wechsler06} for $c(M_{\rm vir})$.

In the previous section we showed how to construct $M_{\rm
  star}-M_{\rm vir}$ relations as a function of redshift.  The
relation between a halo of mass $M_{\rm vir}$ at one epoch and its
mass at some latter epoch is known via Equation \ref{eq:fit2}.  This
relation between halos across time allows us to connect the $M_{\rm
  star}-M_{\rm vir}$ relations across time and hence allows us to
determine the stellar growth of galaxies.

For example, we can start with a halo mass $M_{\rm vir}$ at some early
epoch.  The $M_{\rm star}-M_{\rm vir}$ relation at that epoch then
determines the stellar content of the halo.  We can then evolve this
halo to a later epoch via Equation \ref{eq:fit2}.  With the $M_{\rm
  star}-M_{\rm vir}$ relation at this later epoch we can then read off
the stellar content of the halo at this later epoch.  Continuing this
process allows us to build up the full stellar mass growth of the
galaxy sitting at the center of this evolving halo.  This process can
be repeated for all halos of all masses, allowing one to determine the
stellar mass growth of galaxies as a function of dark matter halo
mass.

Equation \ref{eq:fit2} does not apply to subhalos and yet we have
included subhalos in our analysis up to this point. There are at least
two reasons why this issue will not significantly impact our results.
First, at any given epoch the majority of subhalos were only recently
accreted \citep{Gao04b, Zentner05}, and thus Equation \ref{eq:fit2}
should provide a reasonable approximation to the mass growth history
of subhalos over most of their evolution.  Second, as mentioned above,
subhalos constitute a small fraction of the total halo population
($\sim10-30$\% at any epoch) and thus this approximate treatment
should have a small effect on our conclusions.

\subsection{The impact of merging on galactic growth}
\label{sec:merge}

Galaxies can grow in stellar mass by either star formation or by the
cannibalism of smaller galaxies \citep[e.g.][]{Ostriker77}.  Both
processes can in general contribute to the average stellar mass growth
of galaxies.  We are interested primarily in the inferred star
formation rates as a function of redshift, stellar mass, and halo
mass, and we thus seek a simple way of accounting for the impact of
galaxy merging on the stellar mass growth of galaxies.  In what
follows we present an estimate for the accretion rate of smaller
galaxies onto the halos of larger galaxies.  We then consider two
assumptions for the fates of these accreted systems that will bracket
the range of possibilities.  In the first, we allow all of the
stellar material accreted onto the halo to rapidly merge with the
central galaxy, thereby increasing the mass of the central galaxy.
The other possibility we consider is that the accreted material
remains within the host halo of the central galaxy but does not add to
its measured luminosity.  In other words, the accreted material either
remains as bound satellite galaxies or ends up as diffuse stellar
material not detected in standard survey photometry.  In this latter
scenario stellar mass growth is thus determined entirely by star
formation. These two scenarios are referred to as the ``merger'' and
``no-merger'' scenarios below.  We now describe the merger scenario in
more detail.

Halos grow via the accretion of smaller halos.  The mass spectrum of
accreted halos is approximately self-similar in $m'=m/M_z$ where $m$
and $M_z$ are the mass of the accreted halo and mass of the parent
halo at redshift $z$ \citep{Lacey93, Stewart08}.  The spectrum can be
approximated as:
\noindent
\be\label{eqn:dfdlnm}
\frac{{\rm d}f}{{\rm dln }m'} = \frac{\sqrt{m'}}{2.6}{\rm exp}\bigg[-\bigg(\frac{m'}{0.7}\bigg)^6\bigg],
\ee
\noindent
where $f$ is the fraction of mass accreted in clumps of mass $m'$.
The exponential cut-off is steep because $m'>1$ is not allowed by
definition.  Equation \ref{eqn:dfdlnm} is a fit to the simulation
results of \citet{Stewart08}.  This function does not integrate to
unity, indicating that a significant fraction, $\sim30-50$\%, of the
parent mass is accreted in a diffuse component of dark matter
\citep{Stewart08}.  Whether or not this component is truly diffuse or
is in clumps of very small mass \citep[e.g.][]{Madau08} is immaterial
for our purposes, because in either case this component will not bring
in additional stars.

With the mass accretion spectrum in hand, the halo growth rate,
$\dot{m}_{\rm halo}$, can be converted into a stellar growth rate due
to mergers, $\dot{m}_{\rm stars}$, via:
\noindent
\be\label{eqn:merger}
\dot{m}_{\rm stars} = \dot{m}_{\rm halo} \int \frac{{\rm d}f}{{\rm dln }m'}\frac{M_{\rm star}}{M_{\rm vir}}(M_{\rm vir},z) {\rm d ln}m',
\ee
\noindent
where $\frac{M_{\rm star}}{M_{\rm vir}}(M_{\rm vir},z)$ is the
redshift- and halo mass-dependent stellar-to-halo mass ratio
determined in $\S$\ref{sec:am}.  Equation \ref{eqn:merger} can be
thought of as a convolution of the halo mass accretion spectrum with
the relations between halo mass and stellar mass determined in
previous sections.  In other words, Equation \ref{eqn:dfdlnm} tells us
the types of halos that are accreted, and $\S$\ref{sec:am} tells us
the stellar content of these accreted halos. 

In the following sections we will present results for both the merger
and no-merger scenarios --- two scenarios that bracket reality.  We
remind the reader that while we discuss these scenarios separately, in
reality there is clear evidence that both cases occur.  In particular,
there exists direct observational evidence for galaxy mergers, and the
existence of gravitationally bound groups and clusters of galaxies
implies that not all galaxies merge when their halos merge.  Thus,
when in later sections we discuss a preference for one scenario over
another, we do not mean to suggest that the other scenario never
occurs but rather that it is of sub-dominant importance when
attempting to describe the statistical properties of galaxies.

In $\S$\ref{s:whysfr} we discuss how our results bear on the relative
importance of merging and star formation on the stellar growth of
galaxies.

\subsection{From galactic growth to star-formation rates}
\label{sec:sfr}

With an estimate for the amount of stellar mass growth that is due to
merging/accretion, the SFR of an average galaxy can then be estimated
straightforwardly via a derivative of the portion of stellar mass
growth attributed to star formation.  The relation between mass growth
and star formation is complicated by mass loss due to dying stars.  We
take into account this effect with the following formula:
\noindent
\be
f_{\rm loss}(t) = 5\times 10^{-2}\, {\rm ln}\bigg(\frac{t+3\times 10^5}{3\times 10^5}\bigg),
\ee
\noindent
where $t$ is in years and $f_{\rm loss}(t)$ is the fraction of mass
lost by time $t$ for a co-eval set of stars.  This formula is a fit to
the mass loss of simple stellar populations with a \citet{Chabrier03}
IMF \citep{Renzini93, Bruzual03}.  Note that only $\sim60$\% of the
stellar mass formed in a burst of star formation remains after several
gigayears, and that the stellar mass remaining includes stellar
remnants.  With the full stellar mass growth curve one can then
iteratively solve for the star formation rate required to generate
such growth given the above mass-loss rate formula.

As mentioned in $\S$\ref{sec:gsmf}, the form we have chosen for the
redshift evolution of the stellar mass function is not unique.  Yet it
is clearly the rate of evolution of the mass function that determines
the resulting star formation rates of galaxies in our model.  We
re-emphasize that this form was chosen to best match the observed
cosmic star formation rate density.  Our goal is not to find a unique
form for the evolution of the mass function but rather to present a
consistent framework in which to interpret a vast array of
observational data, and to link that data to the underlying dark
matter skeleton.

\section{Model Implications}
\label{sec:imp}

The previous section presented a method for connecting the stellar
masses and star formation rates of galaxies to dark matter halos over
a range of epochs.  This section explores these relations and compares
to observations where possible.

\subsection{Halo-galaxy connections}

\begin{figure}[!t]
\plotone{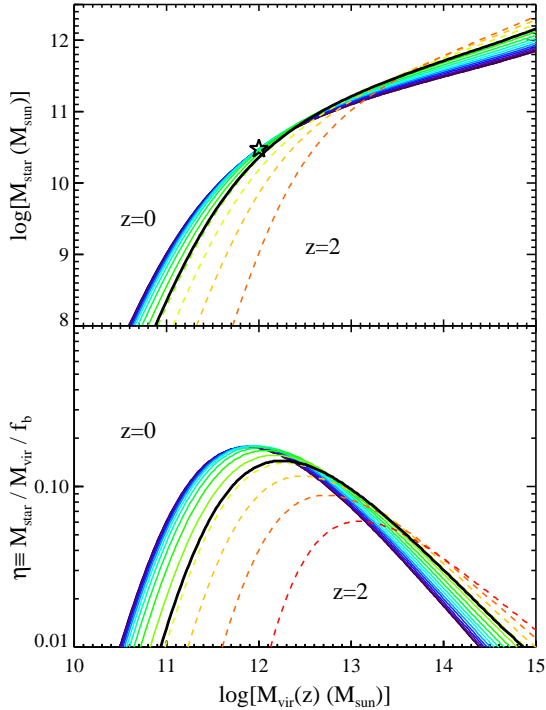}
\vspace{0.5cm}
\caption{{\em Top Panel:} The relation between galaxy stellar mass and
  halo mass from $z=2$ to $z=0$, using the abundance matching model.
  {\em Bottom Panel:} Fraction of available baryons that have turned
  into stars (integrated star formation efficiency) as a function of
  the halo mass and redshift, where $f_b$ is the universal baryon
  fraction.  The star marks the location of the Milky Way at $z=0$.
  The thick black line represents the relation at $z=1$.  The
  relations at $z>1$ ({\it dashed lines}) should be treated with
  caution; see $\S$\ref{sec:highz} for details.}
\label{fig:theplot}
\vspace{0.5cm}
\end{figure}

Implementing the abundance matching technique discussed in
$\S$\ref{sec:am} at various epochs yields the relations between halo
and galaxy mass shown in the top panel of Figure \ref{fig:theplot}.
The generic shape of the relation is governed simply by the
Schechter-like functions of both the galaxy stellar MF and the halo
MF.  The redshift outputs are spaced equally in $(1+z)^{-1}$.  One
novel conclusion drawn from this figure is that, since $z\sim2$, the
stellar mass of galaxies residing in halos of mass $\sim10^{12.5}
\Msun$ stays roughly constant, at $\sim 10^{11} \Msun$.  Over the
redshift range considered, above this mass scale, halo growth
out-paces stellar mass growth, while below this scale, galaxy growth
is more vigorous than halo growth.

The relation shown in Figure \ref{fig:theplot} applies to central
galaxies and to satellites where the halo mass refers to the mass at
the time of accretion onto their host \citep[see, e.g.][ for a
discussion]{Conroy06a}.  Including a modest level of scatter between
stellar and halo mass (as discussed in $\S$\ref{sec:am}) does not
substantially impact the mean relation shown in this plot.  As a
comparative reference, we show the location of the Milky Way in this
figure, as determined from the halo mass estimates of
\citet{Klypin02}.  The Milky Way falls directly on our mean relation.

The shape and mild evolution of the $M_{\rm star}-M_{\rm vir}$
relation shown in Figure \ref{fig:theplot} provides a clear
interpretation of the observed relation between stellar and halo mass
from $z\sim1$ to $z\sim0$ reported in \citet{Conroy07a}.  These
authors used the dynamics of satellite galaxies orbiting around
brighter host galaxies to constrain halo masses, and found that in
bins of galaxy stellar mass, halo mass evolves little or not at all
since $z\sim1$ below $M_{\rm star}<10^{11} M_\Sun$ but increases by a
factor of several above this stellar mass.  This qualitative trend is
evident in Figure \ref{fig:theplot}, and can be attributed to the fact
that above $M_{\rm star}\sim10^{11} M_\Sun$ the relation shallows,
implying that a small shift in the relation over time produces a large
change in the halo mass of a given galaxy mass over time.  

Using a similar approach to our abundance matching technique,
\citet{Shankar06} find comparable results on the evolution of the
$M_{\rm star}-M_{\rm vir}$ relation since $z\sim1$.  Moreover, group
catalogs constructed from large observational surveys have begun to
probe the $M_{\rm star}-M_{\rm vir}$ relation at $z\sim0$
\citep{Berlind06, Yang07}.  Results from the catalogs are in good
agreement with what we find at $z\sim0$.

\begin{figure}[!t]
\plotone{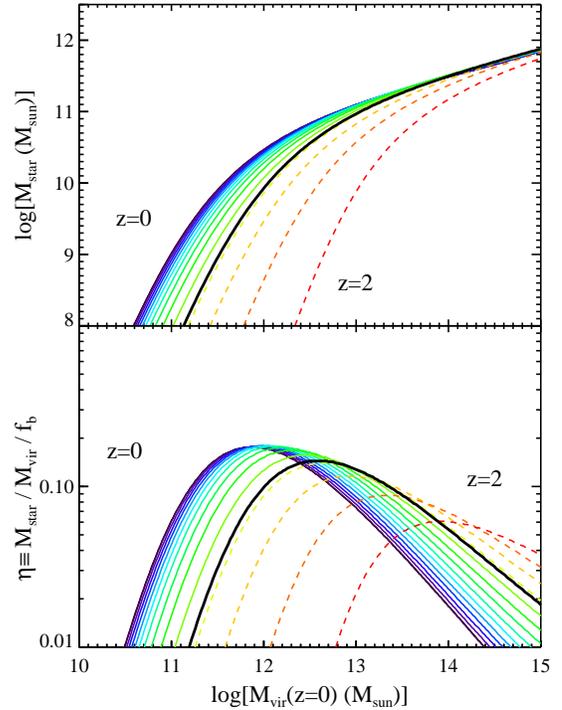}
\vspace{0.5cm}
\caption{{\em Top Panel:} The redshift-dependent relation between
  galaxy stellar mass and the mass of the halo to which it will belong
  at $z=0$.  By focusing on a fixed $z=0$ halo mass, one can read up
  in the plot from $z=2$ to $z=0$ to infer the stellar mass growth of
  an average galaxy that, by $z=0$, resides in that halo.  {\em Bottom
    Panel:} Fraction of available baryons that have turned into stars
  as a function of the $z=0$ halo mass.  The thick black line
  represents the relation at $z=1$.  The relations at $z>1$ ({\it
    dashed lines}) should be treated with caution; see
  $\S$\ref{sec:highz} for details.}
\label{fig:theplot_ev}
\vspace{0.5cm}
\end{figure}

We define the integrated efficiency of star formation as $\eta\equiv
M_{\rm{star}}/M_{\rm{vir}}/f_b$, where $f_b=0.17$ is the universal
baryon fraction \citep{Spergel07}.  This efficiency quantifies the
fraction of available baryons that have been converted into stars, and
peaks where integrated star formation is most efficient.  Abundance
matching readily predicts $\eta(M_{\rm vir})$ and is shown as a
function of redshift in the bottom panel of Figure \ref{fig:theplot}.
The first thing to note is that this result implies that the overall
efficiency of converting baryons into stars is quite low, never
reaching more than $\sim 20\%$ of the potentially available
baryons. Although perhaps somewhat surprising, note also that the
global stellar mass density is $\sim $4--8 times less than the global
baryon density.  This low efficiency is also in good agreement with
current estimates for the total and stellar mass of the Milky Way
\citep{Klypin02}, with estimates of halo masses from weak lensing
measurements combined with stellar mass estimates
\citep{Mandelbaum06}, with halo occupation models both at $z\sim0$ and
$z\sim1$ \citep{Zheng07}, and with estimates from the accounting of
baryons in various states \citep[e.g.][]{Fukugita98, Bell03b,
  Baldry08}.

Note that this quantity is {\it not} the instantaneous star formation
efficiency because $\eta$ herein is defined with respect to the stars
that still exist in the galaxy.  For the IMF we adopt, approximately
40\% of the stellar mass that forms is rapidly lost as massive stars
die.  One may convert $\eta$ into the fraction of baryons that have
ever spent time in a star by dividing the numbers we quote by 0.6.
Thus, for example, the peak shown in the bottom panel of Figure
\ref{fig:theplot} would be $\sim 30\%$ if one were interested in the
instantaneous star formation efficiency.

Two important trends are apparent.  First, the location of the peak
decreases to lower halo masses with time.  The latter trend is a
manifestation of at least one meaning of ``downsizing''
\citep{Cowie96}, and is a natural implication of the fact that the
characteristic stellar mass evolves more slowly than the
characteristic halo mass (see $\S$\ref{s:down} for a discussion of
downsizing).  Second, the amplitude of the peak star-formation
efficiency increases with decreasing redshift, although the magnitude
of this trend is somewhat uncertain.  This is due to the rather
uncertain evolution of $\phi^*$, which directly affects the evolution
of the peak of $\eta(M_{\rm vir})$; varying the evolution in $\phi^*$
over a reasonable range changes the amount of evolution in the peak by
less than a factor of two.  More accurate observational constraints on
$\phi^*(z)$ are required to more robustly pin down the evolution in
$\eta(M_{\rm vir})$.  However, the trend that the peak shifts to lower
masses with time is robust to uncertainties in $\phi^*(z)$.  The mass
at which baryons are most efficiently converted into stars shifts by
about a factor of $\sim20$ from $z=2$ to $z=0$, from $M_{\rm{vir}}
\sim 10^{13}\,\Msun$ to $M_{\rm{vir}} \sim 10^{11.7} \,\Msun$.

These trends, including the factor of $\sim2$ increase in peak
efficiency from $z\sim1$ to $z\sim0$, and the mild decrease in halo
mass at which the peak occurs, agree well with halo occupation
modeling of galaxies at these epochs \citep{Zheng07}.

As discussed in $\S$\ref{sec:mah}, we know from $N$-body simulations
the full mass growth histories of dark matter halos statistically for
a given cosmology. These accretion histories allow us to evolve a halo
of mass $M_{\rm vir}(z)$ at redshift $z$ forward in time to the mass
such a halo will have by $z=0$, $M_{\rm vir}(z=0)$.  We can couple
these halo accretion histories to the $M_{\rm star}-M_{\rm vir}$
relations discussed above to determine the relation between a halo's
mass at $z=0$ and the stellar mass content of that halo as a function
of redshift.

The resulting relations are shown in the top panel of Figure
\ref{fig:theplot_ev}.  The $z=0$ halo mass can be thought of as a
unique tag for each (average) galaxy.  A vertical slice through Figure
\ref{fig:theplot_ev} thus traces out the trajectory of an average
galaxy at different redshifts.  From this plot one can thus read off
\emph{directly} the average stellar mass build-up of galaxies as a
function of their $z=0$ halo mass.  These relations are uniquely
determined by the relations between galaxies and halos at fixed epochs
in conjunction with the evolution of halos demanded by our fiducial
cosmology.

\begin{figure}[!t]
\plotone{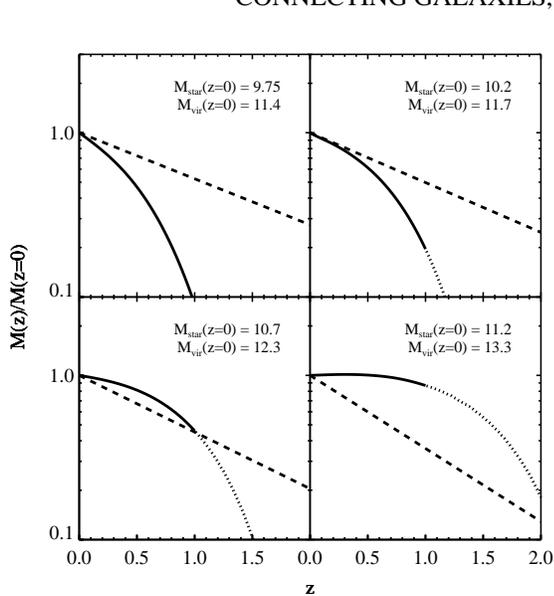}
\vspace{0.5cm}
\caption{Fraction of mass assembled as a function of redshift.  Each
  panel compares the fraction of a galaxy's stellar mass that is
  assembled ({\it solid and dotted lines}) to the fraction of the
  galaxy's parent halo mass that is assembled ({\it dashed lines}).
  The stellar growth curves are dotted at $z>1$ to indicate that this
  regime should be treated with caution; see $\S$\ref{sec:highz} for
  details. The four panels display the assembly history for a range of
  galaxies with $z=0$ stellar and halo masses shown in the legend, in
  units of ${\rm log}(\Msun)$.  It is clear that, at lower mass, a
  larger fraction of the halo is in place at early times compared to
  higher mass.  The opposite trend is true for stellar masses.}
\label{fig:star_growth}
\vspace{0.5cm}
\end{figure}

A clear and robust inference from this figure is that the stellar mass
of galaxies residing in $z=0$ halos of mass $\gtrsim10^{14} \Msun$ was
mostly assembled by $z\sim2$.  This agrees qualitatively with the
modest evolution in the massive end of the observed galaxy stellar
mass function since $z\sim2$ \citep[e.g.][]{Fontana04, Drory04,
  Bundy05, Borch06, Fontana06, Cimatti06, Andreon06, Brown07, Brown08,
  Perez-Gonzalez08, Cool08}.  Note however that the input to our model
is the observed galaxy stellar mass function, so for example if
observations do not account for the low surface brightness
intracluster light associated with central galaxies in massive halos
\citep[e.g.][]{Gonzalez05, Zibetti05, Krick07}, then our model will
also fail to incorporate this component. The diffuse light could
contain as much mass as the central galaxy, and should thus be taken
into account when modeling massive galaxies \citep[see discussion
in][]{Monaco06, Conroy07b, Conroy07c, Purcell07}.  Here however our
focus is on stellar growth and star formation in more modestly-sized
halos.

The bottom panel of Figure \ref{fig:theplot_ev} shows the integrated
star formation efficiency of galaxies at various epochs as a function
of their $z=0$ halo mass.  For halo masses less than
$\sim10^{12}\,\Msun$ the integrated efficiency is a monotonically
increasing function of time.  At higher masses the efficiency rises,
peaks, and then falls with increasing time, and at masses greater than
$\sim10^{14}\,\Msun$ the integrated efficiency is a continually
decreasing function of time since $z=2$.

\begin{figure}[!t]
\plotone{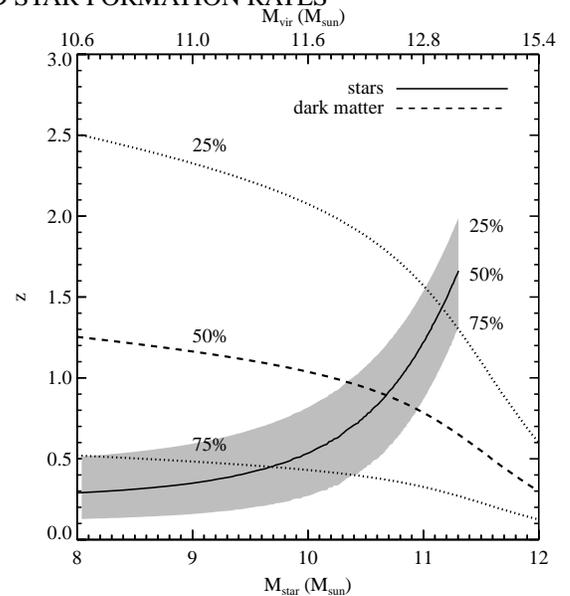}
\vspace{0.5cm}
\caption{Average formation times for galaxies ({\it solid line and
    shaded region}) and halos ({\it dashed and dotted lines}) as a
  function of $z=0$ galaxy and halo mass.  Lines and shaded region
  indicate the redshift at which 25\%, 50\%, and 75\% of the final
  mass was assembled, as labeled in the figure.  For example, a galaxy
  with stellar mass $10^{10} \Msun$ at $z=0$ assembled 50\% of its
  final mass by $z\approx0.5$.  Such a galaxy resides in a halo of
  mass $10^{11.6}\Msun$, which was half assembled by $z\approx1.0$.
  At $M_{\rm star}\lesssim10^{10.7}\Msun$ halos are assembled before
  galaxies while at higher masses the opposite is true, in agreement
  with the trends seen in Figure \ref{fig:star_growth}.}
\label{fig:zhalf}
\vspace{0.5cm}
\end{figure}

\subsubsection{Galaxy growth versus halo growth}

A perhaps more revealing illustration of the results in Figure
\ref{fig:theplot_ev} are shown in Figure \ref{fig:star_growth}.  There
the stellar mass growth of a galaxy is compared to the growth of its
parent dark matter halo for four representative $z=0$ stellar masses.
For stellar masses $\lesssim10^{10.7}\Msun$ (corresponding to $M_{\rm
  vir}\lesssim10^{12.3}\Msun$), fractional halo growth since $z=2$ is
much more mild than stellar growth.  This low-mass regime can thus be
thought of as `internally-dominated', where growth is not controlled by
extra-halo processes.  Stellar growth in this regime is thus driven by
gas physics related to cooling, star formation, and feedback.

The situation is qualitatively different at higher masses.  At stellar
masses $\gtrsim10^{11}\Msun$ (corresponding to $M_{\rm
  vir}\gtrsim10^{13}\Msun$), fractional halo growth is much stronger
than stellar growth at $z<2$.  This is not surprising in light of the
fact that the massive end of the stellar mass function appears to be
approximately in place since $z\sim1$ \citep{Fontana04, Drory04,
  Bundy05, Borch06, Fontana06, Cimatti06, Wake06, Brown07, Cool08}.
This high-mass regime is thus `externally-dominated', in contrast to
lower-mass systems.  High-mass systems are primarily accreting copious
amounts of dark matter, some fraction of which will bring in bound
stellar systems (e.g. satellite galaxies).  The system --- defined
loosely as the region within the halo virial radius --- is thus
growing in a larger sense, while the galaxy at the center of the halo
is not.  This figure does not include the potentially massive
component of stellar light associated with a diffuse background, known
as the intracluster light.  For a discussion of the importance of this
component see \citet{Conroy07b} and references therein.

The average formation times for galaxies and halos is shown in Figure
\ref{fig:zhalf}.  Here we plot the redshift at which 25\%, 50\%, and
75\% of the final mass was assembled, for both the stars accreted onto
the galaxy and for the dark matter mass accreted onto the halo.  Since
we are investigating only average properties, recall that in our model
each galaxy mass is assigned a unique halo mass, and thus each average
galaxy has a unique halo mass and stellar mass, and unique formation
times for both of those components.  It is clear that the stellar mass
in lower mass systems formed later than in high mass systems, while
the dark matter halos display the opposite trend.  Although this
general trend has now been evident from a range of data, this figure
ties together the available information on galaxy and halo growth.

Notice that the results presented thus far do not require making any
assumption for {\it how} the galaxy mass is built up (i.e. whether by
star-formation or merging).  These results only require knowledge of
the redshift-dependent galaxy-halo connections in conjunction with how
average halos grow with time.  We now turn to the more challenging
task of constraining the possible modes of galaxy growth.

\begin{figure}[!t]
\plotone{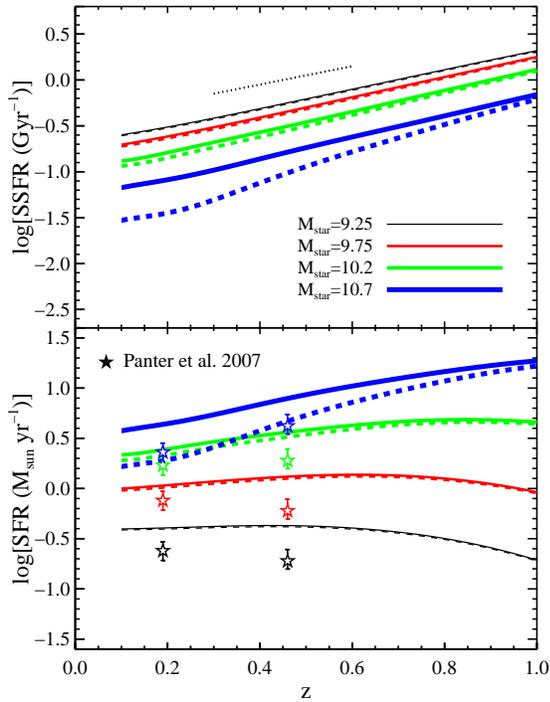}
\vspace{0.5cm}
\caption{ {\em Top Panel:} Average specific star formation rate
  history as a function of redshift, for galaxies with $z=0$ stellar
  masses given in the legend (in units of ${\rm log}(\Msun)$).  The
  dotted line indicates a slope of unity.  {\em Bottom Panel:} Star
  formation rate history as a function of redshift for the same model
  galaxies.  The solid and dashed lines represent our model for two
  prescriptions to relate stellar growth to star formation. The solid
  lines are for the no-merger model while the dashed lines are for the
  merger model (see $\S$\ref{sec:merge} for details).  Observed star
  formation histories from \citet{Panter07} for galaxies of the same
  $z=0$ stellar masses as the model predictions are included for
  comparison.  For reasons discussed in the text, the model is likely
  not reliable at $z>1$.  Note that results for both the model and
  data are {\it average} relations.}
\label{fig:sfr_z}
\vspace{0.5cm}
\end{figure}

\subsection{The star formation history of galaxies}
\label{sec:sfhg}

The star formation rate (SFR) and specific star formation rate (SSFR
$\equiv$ SFR/$M_{\rm star}$) for average galaxy trajectories are shown
in Figure \ref{fig:sfr_z} as a function of redshift for several
representative $z=0$ stellar masses.  The SSFR is approximately
self-similar in galaxy mass for $z<1$ and scales with redshift as
${\rm log(SSFR)}\propto z$.  Lower mass galaxies have higher specific
star formation rates at all times.  In this and subsequent figures we
include two different treatments for the importance of merging on
stellar mass growth.  The solid lines represent the assumption that
all stellar mass growth is due to star formation in situ, while the
dashed lines represent the prescription, described in
$\S$\ref{sec:merge}, that includes mergers and accretion.

The SSFR for the most massive galaxies ($M_{\rm star}\gtrsim 10^{11}
\, \Msun$) is the least constrained in our model because the massive
end of the stellar mass function evolves little at $z<2$, and so the
SFR is a derivative of a nearly constant function.  This can be seen
in the top panel of Figure \ref{fig:theplot_ev}, where it is clear
that small uncertainties/changes in the stellar mass or halo mass
scale can cause relatively large uncertainties in the star formation
rates.  When discussing star-formation rates we restrict ourselves to
stellar masses less than $\sim10^{11}\,\Msun$ where the conversion
between stellar mass growth and SFR is most reliable.

The bottom panel of Figure \ref{fig:sfr_z} shows the average SFR of
galaxies as a function of $z=0$ stellar mass.  The results in this
figure are not directly accessible to observations at high redshift
because the observations do not tell us the connection between
galaxies at different epochs.  Without this information, following the
SFH of a particular galaxy across time is not possible with
observations of the distant past.  However, such information is at
least in principle attainable from consideration of the stellar
populations of galaxies in the local universe.  Attempts at inferring
the SFH of galaxies in this way have concluded that the SFR peaked
earlier for more massive galaxies, and for galaxies less massive than
$\sim10^{10}\Msun$ the data are consistent with a constant SFH, at
least since $z\sim1$ \citep{Lee07a, vanZee01, Heavens04, Panter07}.
The model results presented in Figure \ref{fig:sfr_z} is compared to
the results from \citet{Panter07} who have used the stellar
populations of local galaxies to constrain their mass-dependent star
formation histories.  Our model reproduces the general trends well,
though there is a $\sim0.1-0.2$ dex offset between the model and data.

Furthermore, the trends in Figure \ref{fig:sfr_z} shed light on the
phenomenon known as `downsizing' \citep[e.g.][]{Cowie96, Brinchmann00,
  Juneau05} whereby more massive systems formed the bulk of their
stars at earlier epochs compared to less massive systems.  While it is
clear that star formation has peaked earlier in more massive systems,
it is also apparent from the figure that at any epoch, more massive
galaxies have higher star formation rates and smaller specific star
formation rates than less massive galaxies.  We discuss this issue
further in \S \ref{s:down}.

\subsection{SFR dependence on galaxy and halo mass}
\label{sec:sfrmass}

\begin{figure}[!t]
\plotone{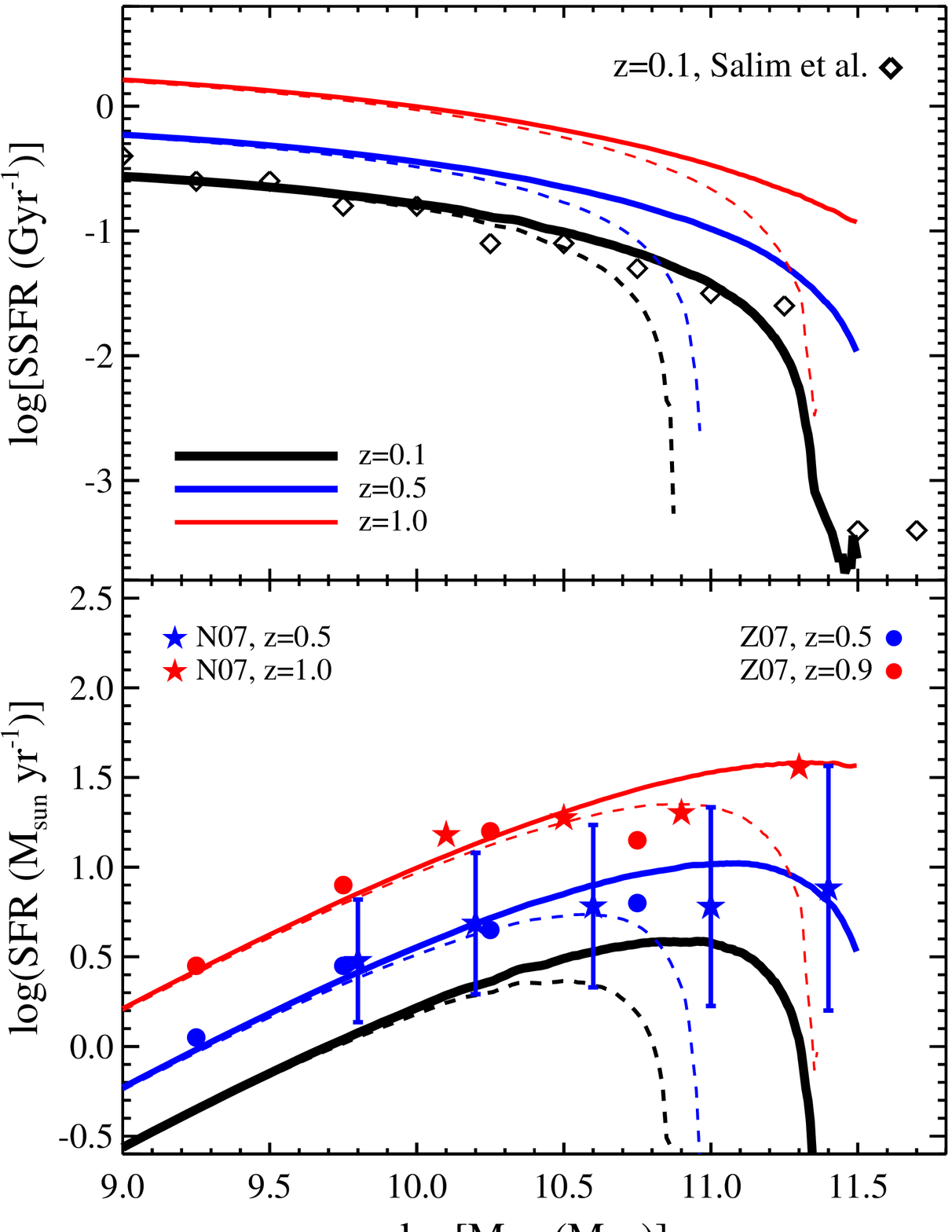}
\vspace{0.5cm}
\caption{SSFR ({\it top panel}) and SFR ({\it bottom panel}) as a
  function of galaxy stellar mass at various epochs.  Note that the
  two panels display equivalent information.  Observational results
  from \citet{Noeske07a} are included and labeled N07, as are $z\sim0$
  data from \citet{Salim07} and data from \citet{ZhengBell07}, labeled
  Z07.  Error bars denote $1\sigma$ scatter, not the error on the
  mean, and are included for only one set of data for clarity. The
  solid and dashed lines represent the no-merger and merger models,
  respectively (see $\S$\ref{sec:merge} for details).  The no-merger
  model is clearly a better match to the data at all epochs.  Note
  that the model describes {\it average} relations of SFR with mass.}
\label{fig:sfr_mass}
\vspace{0.5cm}
\end{figure}

\begin{figure}[!t]
\plotone{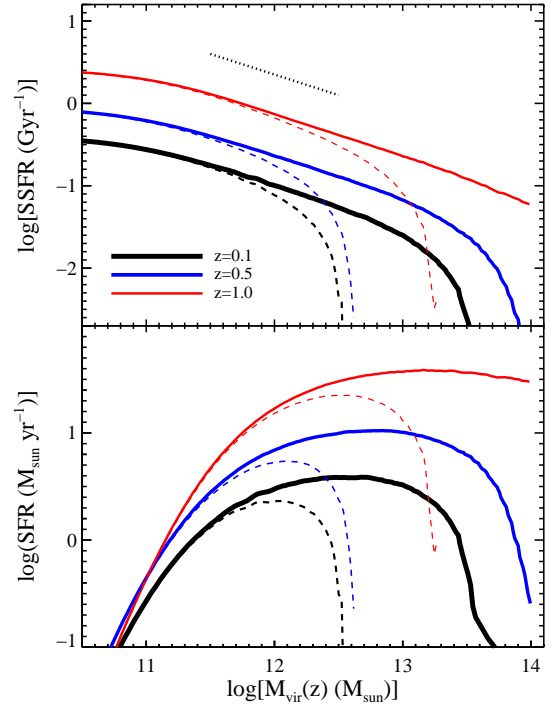}
\vspace{0.5cm}
\caption{SSFR ({\it top panel}) and SFR ({\it bottom panel}) as a
  function of halo mass at various epochs.  As in previous figures,
  the solid and dashed lines represent the no-merger and merger
  models, respectively (see $\S$\ref{sec:merge} for details).  These
  two assumptions only impact the derived SFR at high masses,
  and the data favor the no-merger model (see Figure
  \ref{fig:sfr_mass}).  In the top panel the dotted line indicates a
  slope of $-0.5$.  Note that the model describes {\it average}
  relations of SFR with mass.}
\label{fig:sfr_hmass}
\vspace{0.5cm}
\end{figure}

\begin{figure*}[!t]
\center
\includegraphics[angle=90, width=0.7\textwidth]{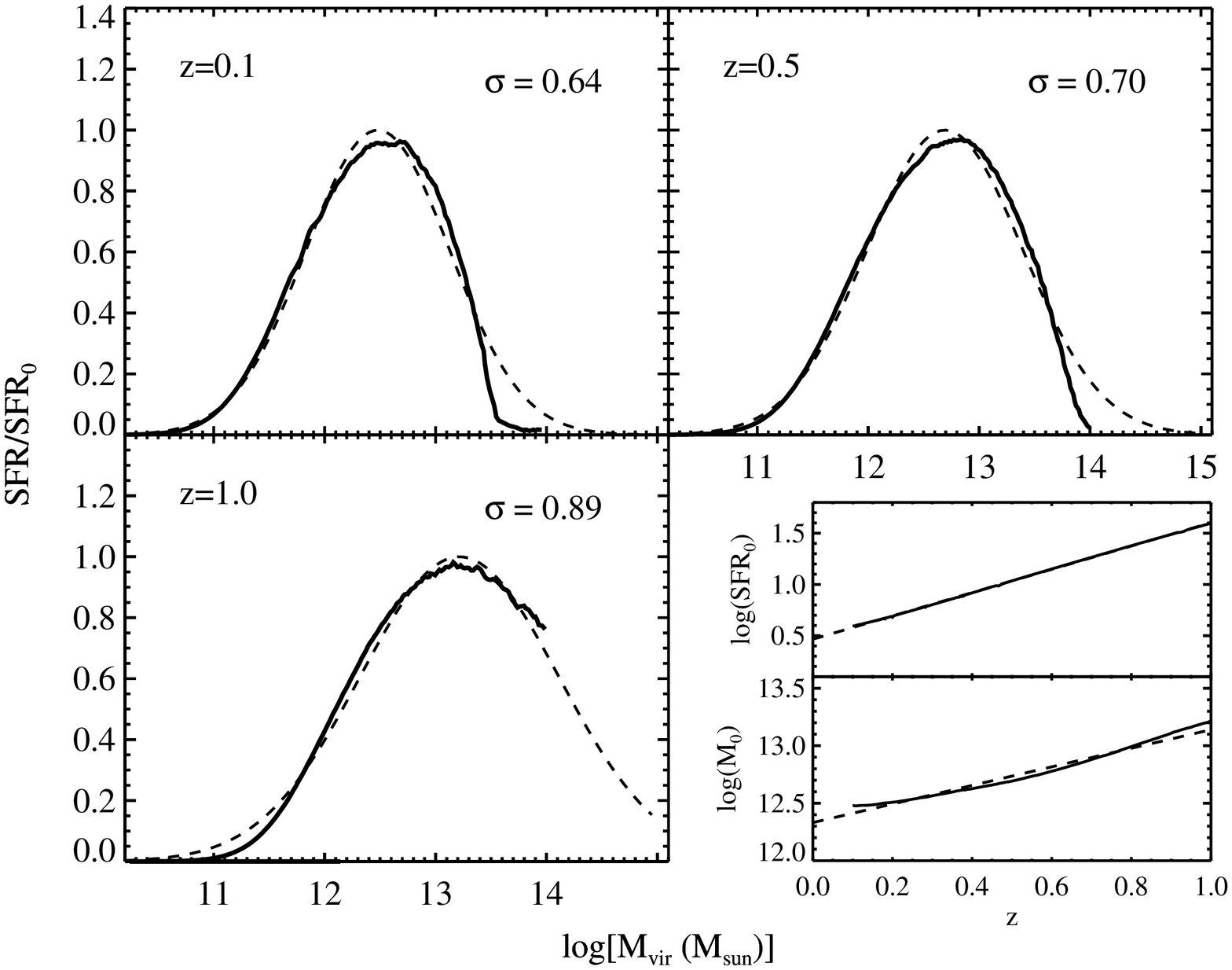}
\vspace{0.5cm}
\caption{SFR as a function of halo mass at various epochs for our
  preferred, no-merger model ({\it solid lines}).  In this figure we
  include Gaussian fits to the SFR($M_{\rm vir}$) relations ({\it
    dashed lines}).  The inset shows the best-fit normalization,
  SFR$_0$, and mean, $M_0$, as a function of redshift ({\it solid
    lines}), along with linear fits ({\it dashed lines}) to these
  relations, which are barely distinguishable from the relations
  themselves.  The dispersion is a weak function of redshift and is
  thus only included in the three larger panels for clarity--- the
average dispersion between $0.1<z<1.0$ is $\sigma=0.72$.  Note that at
  higher redshift the turn-over at high masses is not resolved and the
  fits there should thus be treated with caution.}
\label{fig:sfrfits}
\vspace{0.5cm}
\end{figure*}

Our model allows us to calculate the SSFR and SFR of galaxies as
functions of stellar mass and redshift; these are shown in Figure
\ref{fig:sfr_mass}.  As above, the figure includes two different
prescriptions for relating stellar growth to star formation.  One
possibility is that all stellar growth is due to star formation ({\it
  solid lines}) while the other allows for some fraction of stellar
growth to be attributed to merging ({\it dashed lines}; see
$\S$\ref{sec:merge}).

The model is compared to a variety of data from the literature over
the redshift interval $0<z<1$.  In all cases the data are meant to
represent {\it average} star formation rates as a function of stellar
mass (i.e. the average star formation rate of {\it all} --- both red
and blue --- galaxies at a given stellar mass).  The
completeness-corrected average SFR$-M{\rm star}$ relation from
\citet{Noeske07a} was constructed based on the completeness
corrections of \citet[][K. Noeske private communication]{Lin08} in
order to account for red galaxies with no detectable levels of star
formation.  The results from \citet{ZhengBell07} were derived from stacked
data and can thus be interpreted as average relations.  Finally, the
results from \citet{Salim07} were determined from data with sufficient
sensitivity to detect extremely low levels of star formation and can
thus also be interpreted as an average relation over all galaxies at a
given stellar mass.

It is clear from Figure \ref{fig:sfr_mass} that the assumption that
all stellar growth is due to star formation (i.e. the no-merger
scenario; {\it solid lines}) provides a much better match to the data at
stellar masses $\gtrsim10^{10}\Msun$.  For this reason we adopt this
assumption as the fiducial model.  At lower masses the no-merger and
merger scenarios yield the same predictions for the star formation
rates (i.e., even with maximal merging, incoming satellite galaxies do
not contribute any appreciable stellar mass) The importance of star
formation over merging in galactic growth is discussed further in
$\S$\ref{s:whysfr}.

\begin{figure}[!t]
\plotone{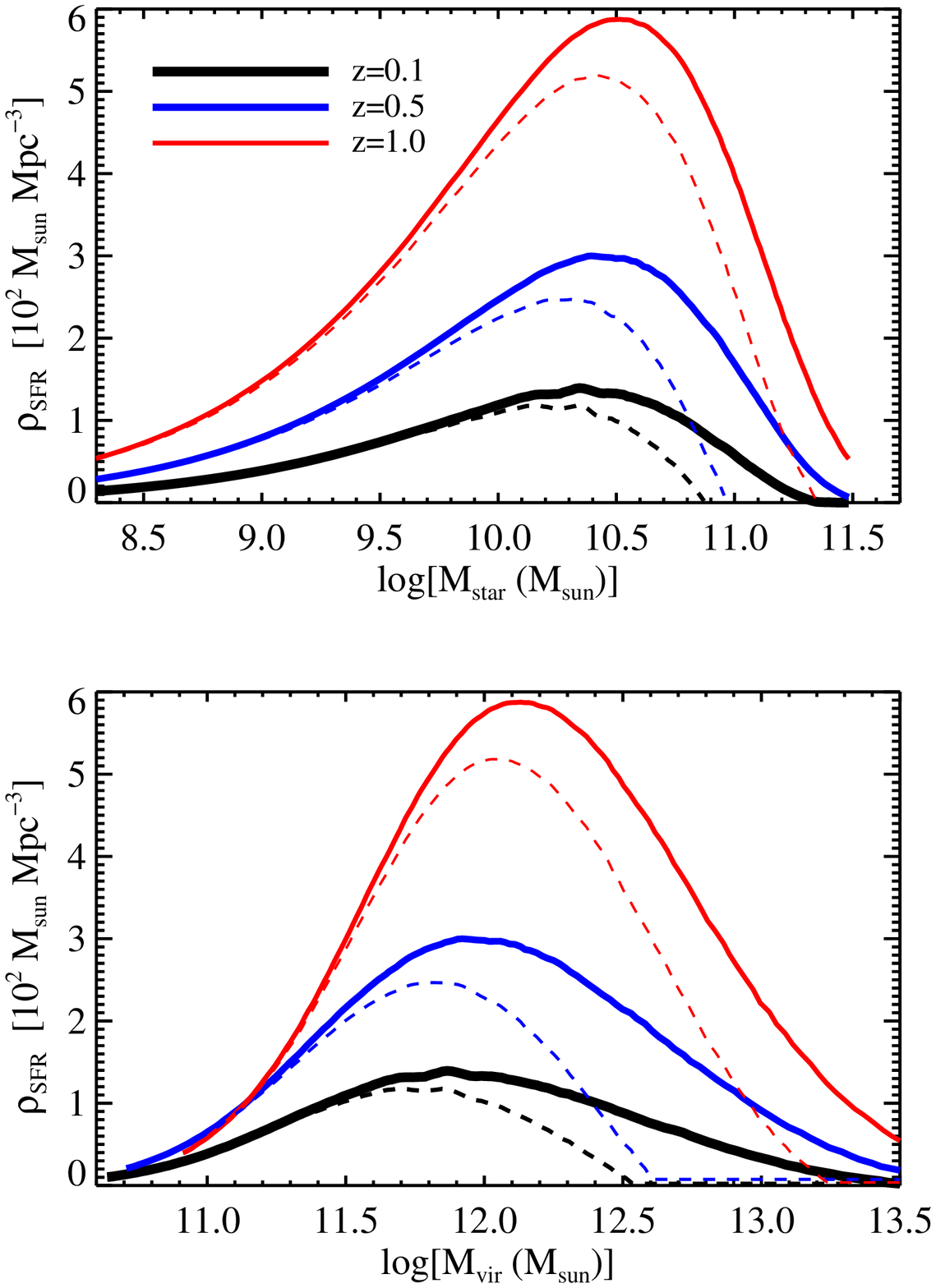}
\vspace{0.5cm}
\caption{{\it Top Panel:} SFR density as a function of stellar mass
  and redshift.  {\it Bottom Panel:} SFR density as a function of halo
  mass and redshift.  These plots illustrate the contribution to the
  global SFR density for galaxies of a given stellar mass ({\it top
    panel}) and for galaxies residing in a given halo mass ({\it
    bottom panel}).  As in previous figures, we include SFR estimates
  for both the assumption that the stellar growth is entirely due to
  star formation ({\it solid lines}), and a simple prescription to
  account for the amount of stellar growth due to mergers and
  accretion ({\it dashed lines}, see $\S$\ref{sec:merge}).  This
  figure demonstrates that the bulk of star formation at $z\leq1$
  occurs in relatively massive galaxies and in halos of mass
  $10^{11.5-12.5}\Msun$. }
\label{fig:rhosfr_zm}
\vspace{0.5cm}
\end{figure}

\begin{figure}[!t]
\plotone{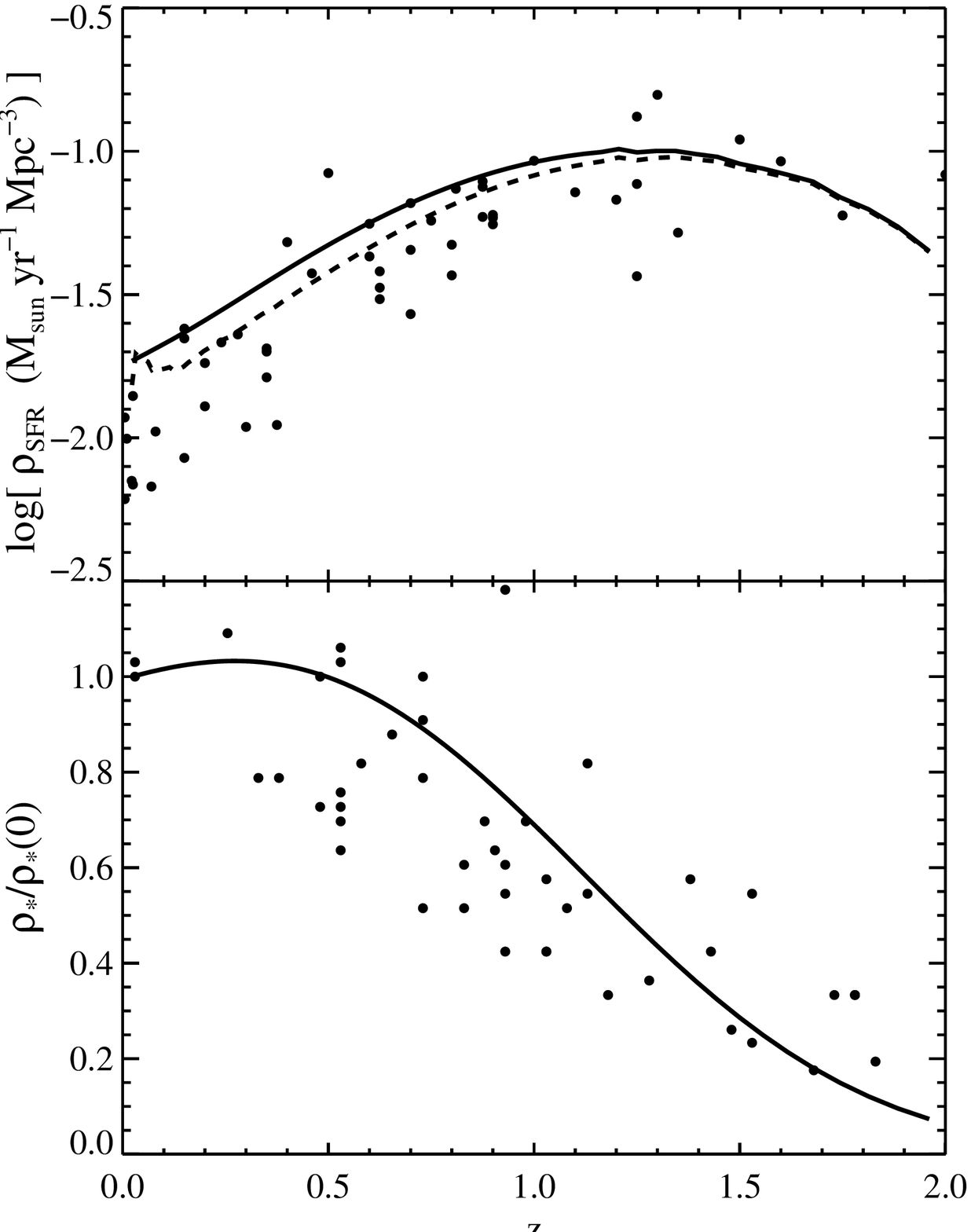}
\vspace{0.5cm}
\caption{\emph{Top Panel}: Cosmic SFR density as a function of
  redshift.  Our model predictions both for the no-merger (\emph{solid
    line}) and merger (\emph{dashed line}) models are compared to the
  data compilation of \citet[][\emph{circles}]{Hopkins04}.
  \emph{Bottom Panel}: Cosmic stellar mass density as a function of
  redshift (\emph{solid line}) compared to the data compilation of
  \citet[][\emph{circles}]{Wilkins08}.}
\label{fig:rhosfr}
\vspace{0.5cm}
\end{figure}

It is instructive to understand which aspects of the model are driving
agreement with the data.  The normalization of the model predictions
depend on the evolution of $\phi^\ast$, the redshift-dependent
normalization of the stellar mass function, while the shapes depends
on $\alpha^\ast$ and $M^\ast$, although the latter two dependencies
are much weaker than the first.  Recall that we have tuned the
evolution of $\phi^\ast$ to reproduce the normalization of the
SFR$-M_{\rm star}$ relations, but not the shape of these relations.
The shape is thus a robust prediction of our approach, while the
normalization agrees with the data by construction.

Figures \ref{fig:sfr_z} and \ref{fig:sfr_mass} can be thought of as
consistency checks between the model and data, since many of the
implications that can be drawn from these figures are readily
available from the data themselves.  In contrast, Figure
\ref{fig:sfr_hmass} contains a variety of novel results.  This figure
shows the model predictions for the SFR and SSFR in galaxies {\em as a
  function of their host dark matter halo masses.}

An interesting consequence of Figure \ref{fig:sfr_hmass} is that the
halo mass at which the most vigorous star formation occurs is not a
strong function of redshift.  In addition, the peak in SFR occurs over
a large range of halo masses, rather than at one well-defined scale.
Moreover, the SSFR$-M_{\rm vir}$ relation appears to be almost
scale-free up to $M_{\rm vir}\sim10^{13}\Msun$ for our favored model,
with the normalization steadily decreasing with time and a
non-evolving slope.  Over the range $10^{11.0}\lesssim M_{\rm
  vir}\lesssim10^{13.0} \Msun$, the redshift- and mass-dependent
relation can be approximated by:
\noindent
\be 
{\rm SSFR}\approx 10^{4.9+0.9z}\, M_{\rm vir}^{-0.5} \,\,\,{\rm Gyr}^{-1}.
\ee

The relation between SFR and halo mass shown in Figure
\ref{fig:sfr_hmass} can be thought of as the most fundamental of the
relations discussed in this work, as this redshift-dependent relation
gives rise to all other relations.  This relation thus provides a
direct link between observations and models in the sense that any
model which reproduces the trends in Figure \ref{fig:sfr_hmass} will
automatically match the variety of observational results discussed
herein.  This connection between SFR and halo mass, determined
entirely from observations with our simple approach, can thus be of
general use to the modeling community in constraining models of
cooling, feedback, and star formation in galaxies.

Figure \ref{fig:sfrfits} shows the SFR$-M_{\rm vir}$ relations again,
now with Gaussian fits (note that the $y$-axis is shown here in linear
units).  It is clear that the relations are well-characterized as
Gaussian except perhaps at the highest masses at low redshift (where
our model is least well-constrained) and at high masses at high
redshift where the turn-over is not resolved.  The fits in these
regimes should thus be treated with caution.

The Gaussian fits are characterized by three parameters: the peak,
SFR$_0$, mean, $M_0$, and dispersion, $\sigma$.  The first two
parameters are shown as a function of redshift in the inset panels of
Figure \ref{fig:sfrfits}.  These parameters are very well approximated
with the following linear relations:
\noindent
\bea
{\rm log(SFR_0)} = 0.47 + 1.1\, z \\
{\rm log}(M_0) = 12.3 + 0.81\, z.
\eea
\noindent
These fits are included in the inset panels.  The dispersion is a much
weaker function of redshift than either the normalization or the mean.
The dispersion ranges from 0.64 at $z=0.1$ to 0.89 at $z=1.0$ with a
mean value of 0.72 over the whole interval $0.1<z<1.0$.  It is
important to recognize that, while the general functional form and
redshift-dependent trends are robust predictions of our model, the
precise values are subject to uncertainty because the observations
themselves, to which the model is tied, still have substantial
uncertainties.

In all of these figures it is important to keep in mind that we are
presenting {\it average} relations between various quantities.  At
first glance Figure \ref{fig:sfr_hmass} might suggest that there
would be no red galaxies (where star-formation has ceased) at
$z\sim1$.  There can of course be such galaxies, as there can also be
galaxies with SFR in excess of the average relation presented in
Figure \ref{fig:sfr_hmass}.

It is worth mentioning here why the approach taken in this paper is
particularly useful.  The current generation of hydrodynamic
simulations and semi-analytic models are not capable of reproducing
the redshift-dependent trends shown in Figure \ref{fig:sfr_mass}
\citep{Dave08}.  The cause of this discrepancy is not currently
understood, although \citet{Dave08} speculates that an evolving IMF
can alleviate the tension.  Regardless, it is clear that until this
tension is resolved, using either hydrodynamic simulations or
semi-analytic models to interpret the observations and connect them to
the formation and evolution of halos requires caution.  Our approach
matches the observations by construction and it can thus be used with
more confidence for interpreting the data.  Its main limitation, and
the main advantage of the simulations, is that our model makes no
reference to the underlying physical processes governing these
relations.  However, the connection between observables and halo mass
derived from our approach should be very helpful in informing these
more physical models.

\subsection{Global properties}
\label{s:glob}

Figure \ref{fig:rhosfr_zm} plots the SFR density as a function of
stellar mass at $z=0.1, 0.5,$ and $1.0$.  This quantity is the SFR
density contributed by galaxies with mass $M_{\rm star}$ and is
produced by multiplying the SFR$-M_{\rm star}$ relation by the galaxy
stellar mass function, $\Phi(M_{\rm star})$.  This quantity is thus
well-constrained by the observational data.  In the figure, this
quantity is plotted both as a function of the stellar and dark matter
halo mass.

In the top panel, the peak for our favored model ({\it solid lines})
is $\sim0.5$ dex lower than the characteristic mass of the stellar
mass function, $M^\ast$, at all epochs, indicating that the bulk of
the SFR density is contributed by galaxies a factor of $\sim3$ in mass
below $M^\ast$.  In other words, the characteristic galactic mass in
which stars form since $z\lesssim1$ is a factor of $\sim3$ lower than
the characteristic galactic mass dominating the mass density.

This peak in the SFR density does not change appreciably from $z\sim0$
to $z\sim1$.  The bulk of star formation at $z\lesssim 1$ never occurs
in small systems, rather it is always dominated by relatively massive,
$M_{\rm star}\sim10^{10.0-10.5}\Msun$ galaxies.  Similarly, the peak
as a function of $M_{\rm vir}$ ({\it bottom panel}) decreases only
slightly, by at most $\sim0.5$ dex from $z=1$ to $z=0$.  These results
imply that a typical star in the Universe forms in galaxies of similar
mass, both in terms of stars and dark matter, from $z\sim1$ to
$z\sim0$.  The principle difference with redshift is that the
distribution of masses increases with time, so that stars are more
likely to form in a variety of systems at later epochs.

Finally, Figure \ref{fig:rhosfr} compares our model predictions for
the evolution of the cosmic SFR density and stellar mass density to
data compilations provided by \citet{Hopkins04} and \citet{Wilkins08},
respectively.  The top panel includes the model prediction for the SFR
under the two different assumptions discussed in $\S$\ref{sec:merge}.
The agreement between model and data at $z<1$ in this figure is
largely by construction, as mentioned in $\S$\ref{sec:gsmf}, but is
included here for completeness.  The discrepancy between model and
data at $z>1$, which is more apparent in Figure \ref{fig:gsmf} but is
also seen here, is discussed in the next section.

\subsection{The model at $z>1$}
\label{sec:highz}

In this section we have focused largely on redshifts less than one.
At higher redshifts the model fails to match the observed cosmic SFR
stellar mass density evolution at $z>1$ as shown in Figure
\ref{fig:rhosfr}, and, relatedly, the normalization of the stellar
mass function at $z=2$, shown in Figure \ref{fig:gsmf}.  This
disagreement arises due to a more generic discrepancy between observed
SFR indicators and stellar mass estimates, as discussed in
$\S$\ref{sec:gsmf}.  As discussed in that section, one possible
explanation is that the IMF evolves with redshift
\citep[e.g.][]{Wilkins08}, although we emphasize that this possibility
is controversial.  Nonetheless, the generic inconsistency between {\it
  observations} implies that our model cannot be robustly applied to
$z>1$.

Moreover, it is plausible that our assumption of a tight correlation
between stellar mass and halo mass breaks down at higher redshift
\citep[cf. discussion in][]{Conroy08a}.  This tight correlation is
strongly supported at $0<z<1$ by the observed stellar mass-dependent
autocorrelation function of galaxies, in the sense that more massive
galaxies are more strongly clustered \citep{Li06, Meneux08}.  This
observational result can be most easily explained if more massive
galaxies reside in more massive halos because halo clustering strength
is a monotonically increasing function of halo mass
\citep[e.g.][]{Zehavi05, Conroy06a}.

This observational trend has not been unambiguously confirmed at $z>2$
\citep{Adelberger05}, except perhaps at the very highest masses
\citep{Quadri07}.  It is clear however that there is strong {\it
  restframe UV luminosity}-dependent clustering at these early epochs
\citep{Adelberger05, Lee06, Ouchi05}.  By analogy with stellar masses
at low redshift, this trend can be understood if UV luminosity, and
hence the star formation rate, is strongly and monotonically
correlated with dark matter halo mass.  If this is the correct
interpretation, then our model must be modified at these early epochs
\citep[see e.g.][]{Conroy08a}.

\subsection{Dependence on cosmological parameters}
\label{s:cosmo}

The halo mass functions and halo mass accretion histories in this
model are dependent on cosmological parameters.  Because it affects
the shape and normalization of the mass function, the normalization of
the power spectrum can have a large affect on our results. The
analytic framework for these halo properties described in
$\S$\ref{sec:hmf} allows us to straightforwardly explore the effect of
cosmological parameters on our results.  Here we just consider the
effect of the normalization of the power spectrum as specified by the
rms fluctuations measured in $8\Mpc h^{-1}$ spheres, $\sigma_8$.

We find that the impact of $\sigma_8$ on our results is imperceptible
for $z=0$ halo masses less than $\sim10^{14}\Msun$.  This is due to
the fact that the dependence of the halo mass function on $\sigma_8$
is much stronger at the massive end.  When considering a change from
our fiducial model with $\sigma_8=0.76$ to a model with
$\sigma_8=0.90$, even at $M_{\rm vir}=10^{14}\Msun$ the difference in
halo abundance is $<0.3$ dex, and at $M_{\rm vir}=10^{13}\Msun$ it is
$<0.1$ dex.  The accretion histories are also a function of
$\sigma_8$, but again the effect is only manifest at high halo masses.
Since the bulk of our results focus on halo masses
$\lesssim10^{14}\Msun$, we conclude that the uncertainty in $\sigma_8$
does not impact our conclusions.

\section{Discussion}

\subsection{Downsizing: what, when, and where}
\label{s:down}

The phenomenon known as ``downsizing'', coined by \citet{Cowie96}, has
received much attention recently, and, perhaps confusingly, has been
attributed to a number of related but nevertheless different
phenomena.  In its most general sense the term highlights a shift in a
{\it preferred mass scale} of a phenomenon related to stellar growth
or star formation.  With an observationally-constrained model for the
redshift-dependent connections between star formation, stellar mass,
and halo mass, we are in a position to clarify and outline the
relations between the various meanings of downsizing.  For clarity, we
focus discussion on galaxies with $M_{\rm star}<10^{11}\Msun$ and
$z<1$, where our results are most reliable \citep[see e.g.][for a
theoretical discussion of downsizing for higher mass
systems]{Cattaneo08}.

Originally, downsizing described the observation that the maximum
$K$-band luminosity of galaxies above a SSFR threshold decreases with
time \citep{Cowie96}.  In this definition, the SSFR threshold is
independent of redshift.  In Figure \ref{fig:sfr_mass} we can see that
a line of constant SSFR will indeed include more massive galaxies at
earlier epochs, consistent with this notion.  However, inspection of
the full relations in this figure shows that this notion of downsizing
is driven by the global phenomenon that all galaxies have lower star
formation rates at later times.  In fact, these relations do not
appear to show any preferred scale with stellar mass (except possibly
at very high stellar masses), but rather they shift self-similarly in
time as noted by \citet{Noeske07a, Noeske07b}.

Figures \ref{fig:sfr_hmass} and \ref{fig:sfrfits} display another type
of downsizing in the sense that the dark matter halo mass at which
star formation is most intense shifts to lower masses at later times.
This trend is apparent in both our favored model where stellar growth
is entirely due to star formation ({\it solid lines}) and in the model
that includes stellar growth due to mergers ({\it dashed lines}),
suggesting that this form of downsizing is a generic feature of dark
matter halos.  Indeed, \citet{Neistein06} has argued that downsizing
arises naturally from the accretion histories of the dark matter halos
themselves.  While this is an intriguing possibility, the uncertain
relation between halos and galaxies (connected in their terms by the
competition between gas heating and cooling), makes their conclusions
difficult to interpret at face value.

Downsizing has also been attributed to the observation that more
massive galaxies seem to have formed the bulk of their stars earlier.
This type of downsizing has been referred to as ``archaeological
downsizing'' because it is observed in the fossil record of the
spectra of $z\approx0$ galaxies.  It has been most convincingly
demonstrated in local elliptical galaxies where one finds that more
massive galaxies formed the bulk of their stars earlier and over
shorter timescales than less massive galaxies \citep{Thomas05}.

This form of downsizing can be seen clearly in Figure
\ref{fig:star_growth}, which shows that the most massive galaxies
formed the bulk of their stars earlier than less massive galaxies.
From Figure \ref{fig:sfr_z} it is also clear that the peak of the SFR
occurs at earlier times for more massive systems.  Note however that
in the top panel of Figure \ref{fig:sfr_z} there is no clear scale in
the SSFR except perhaps for the most massive galaxies, and thus there is
little evidence for any type of downsizing in this relation.

A final meaning of downsizing concerns the sites where the bulk of
stars are being formed at any epoch.  This form of downsizing implies
that stars are being formed in preferentially smaller systems at later
times.  In Figure \ref{fig:rhosfr_zm} it is clear however that the
typical masses hosting the bulk of star formation has not changed
appreciably since $z\sim1$.  There is thus no evidence for this form
of downsizing given the available data.

It is understandable, in light of the preceding discussion, that the
term downsizing has been used to describe so many related but
different phenomena, and that some authors find no evidence for a
downsizing phenomenon.  As we have seen in the various relations
between SFR, SSFR, stellar mass, halo mass, and redshift, some show a
shift in preferred scales with time, and some do not.  Downsizing, of
whatever type, thus manifests itself only in certain relations, and
not in others.

\subsection{A characteristic halo mass?}
\label{s:cmass}

Recently several theoretical studies have raised the possibility of a
characteristic halo mass below which star formation occurs, and above
which star formation is truncated \citep{Keres05, Dekel06, Birnboim07,
  Cattaneo07}.  In this section we focus on the observational evidence
for or against a {\it sharp or narrow range in halo masses} over which
galaxy properties, such as star formation rates, change dramatically.

This characteristic mass scale, which is thought to be $M_{\rm
  vir}\sim10^{12}\Msun$ at $z\sim0$, may be related to the observed
stellar mass scale at which many properties of galaxies qualitatively
change \citep[e.g.][]{Kauffmann03b}.  The fundamental gas dynamical
effect occurring in halos above this characteristic scale is thought
to be the formation of a stable shock through which infalling gas must
cross, thereby raising the temperature of this newly accreted gas to
the virial temperature of the halo \citep{Keres05}.  Accreted gas that
is shock-heated is known as `hot-mode' accretion, while gas that is
not shock-heated is referred to as `cold-mode' accretion.

There are a number of outstanding issues related to any possible sharp
transition in e.g. galaxy colors or star formation rates occurring at
a characteristic halo mass.  First, the transition from cold to
hot-mode accretion does not appear to be particularly sharp in
hydrodynamic simulations \citep{Keres05}.  There is clearly a
transition region, but it is broad, spanning the mass range
$\sim10^{11-12}\Msun$.  Moreover, the establishment of a hot
atmosphere does not guarantee that star formation will cease because
such gas will still radiate and can thus cool (although hot, low
density gas is more susceptible to further heating processes than
cool, dense gas).  Indeed, the cooling time of the intracluster medium
at the centers of massive clusters is in many cases $<10^9$ Gyr
\citep{Sanderson06}.  One thus must propose additional mechanisms that
are capable of supplying sufficient energy to keep the hot atmosphere
from cooling and hence forming stars.  Possible mechanisms include
feedback from active galactic nuclei \citep[e.g.][]{Croton06,
  Cattaneo08}, heating by dynamical friction \citep{Miller86,
  Khochfar08}, thermal conduction \citep{Zakamska03}, virialization
heating \citep{Wang08}, heating by ram-pressure drag \citep{Dekel08},
and supernovae heating.  The relevance of these or other mechanisms to
the shut-down of star formation is currently a subject of active
debate \citep[see][for a recent evaluation]{Conroy08b}.

Our results on the relation between star formation rates and halo
masses can shed light on this issue.  In particular, our model
provides a bridge between the observations and the underlying dark
matter structure.  At stellar masses $\lesssim10^{11}\Msun$, where our
results are most reliable, we find no significant evidence for a {\it
  sharp} characteristic halo mass at which star formation rates
dramatically change, when considering average relations between star
formation, stellar and halo mass.  This statement is based on the
following inferences.

The scale at which galaxy properties such as color and morphology
appear to change qualitatively is at $M_{\rm star}\sim10^{10.3}\Msun$
\citep{Kauffmann03b}.  Our abundance matching results shown in Figure
\ref{fig:theplot} demonstrate that this stellar mass corresponds to a
halo mass of $\sim10^{12}\Msun$ at $z\sim0$, in qualitative agreement
with the characteristic halo mass scale mentioned above
\citep{Dekel06}.

In Figure \ref{fig:sfr_hmass}, however, for our favored model (the no
merging model) there is no abrupt change in the average star formation
rate as a function of halo mass for $M_{\rm vir}\lesssim10^{13}\Msun$,
at either $z\sim0$ or at higher redshifts.  Instead, over this range
the SFR$-M_{\rm vir}$ relation is approximately Gaussian with a broad
peak at $M_{\rm vir}\lesssim10^{12.5}\Msun$ at $z\sim0$.  We reiterate
that the $z\sim0$ SFR$-M_{\rm vir}$ relation is determined by 1) the
$z\sim0$ SFR$-M_{\rm star}$ relation, where the model and data agree
well, and 2) our connection between stellar and halo mass at $z\sim0$.
This latter connection is known to reproduce the observed clustering
properties of galaxies \citep{Conroy06a, Zheng07} and also agrees with
weak lensing measurements of halo masses as a function of stellar mass
\citep{Mandelbaum06}.  At higher redshifts the peak shifts to higher
halo masses, and the gradual roll-over at the high-mass end seen in
the $z\sim0$ average SFR$-M_{\rm vir}$ relation disappears.  The data
is thus consistent with there being no drop in star formation
whatsoever above a given halo mass scale at higher redshifts, at least
for halos with mass $M_{\rm vir}\lesssim10^{13}\Msun$, where we focus
our results.  In this figure the merger model does indeed produce a
sharp break in the SFR$-M_{\rm vir}$ relation, but recall that this
model fails to reproduce the observed SFR$-M_{\rm star}$ relation
shown in Figure \ref{fig:sfr_mass}.

It is important to stress that these conclusions at $z\sim0$ rest on
the reliability of the observed $z\sim0$ average SFR$-M_{\rm star}$
relation, as reported by \citet{Salim07} \citep[see
also][]{Schiminovich07}.  These authors caution that the star
formation rates inferred for massive galaxies, $M_{\rm
  star}\gtrsim10^{11.5}\Msun$, may in some cases be upper limits
because low levels of $UV$ flux may arise from old stellar populations
\citep[e.g.][]{Rich05} that are not typically included in modeling of
star formation rates.  Interpreting low levels of $UV$ flux has
historically been challenging for this reason.  Similar issues arise
at higher redshifts.  Note however that we do not rely on these
massive galaxies for our conclusions because they reside in very
massive, $M_{\rm vir}>10^{14}\Msun$, halos.

Moreover, as shown in Figure \ref{fig:sfr_hmass}, there is clearly no
scale in the specific star formation rate as a function of halo mass
--- it is approximately a power law that scales as SSFR$\propto M_{\rm
  vir}^{-0.5}$ over at least two orders of magnitude in halo mass.
Again, these statements apply to halos with mass $M_{\rm
  vir}\lesssim10^{13}\Msun$.  At higher masses our model is not
well-calibrated.

In sum, while a well-defined characteristic halo mass, above which
star formation is truncated, may be an appealing mechanism for
generating red sequence galaxies \citep{Cattaneo08}, there is no clear
indication from our data-driven model that this scale is particularly
sharp.  It is clear that observed galaxy properties change
qualitatively around a stellar mass scale of $M_{\rm
  star}\sim10^{10.3}\Msun$, corresponding in our model to a halo mass
of $\sim10^{12}\Msun$.  We simply emphasize that the data favors a
rather gradual shift in galaxy properties across this halo mass scale.

\subsection{The relative importance of star formation and merging to
  galactic growth}
\label{s:whysfr}

In order to translate our model predictions for stellar mass growth
into predictions for star formation rates, we have to make assumptions
for the fraction of mass growth attributed to mergers, as described in
$\S$\ref{sec:merge}.  One approach is to assume that all stellar
material accreted onto the halo remains in the halo as satellite
galaxies or is stripped and remains in the stellar halo.  This is the
no-merger model described above.  In this model all stellar growth is
due to star formation. The second approach is to assume that all of
the accreted material immediately falls onto the central galaxy and
hence contributes to its stellar growth.  These two approaches should
bracket the range of possibilities, as in reality some accreted
material will lose energy and merge with the central galaxy, while
other material will remain as bound satellites, or will merge with the
central galaxy yet be dispersed outside the photometric radius.

For galaxies with stellar mass $\lesssim10^{10}\Msun$, these two
treatments for the importance of merging on stellar growth lead to
indistinguishable predictions for the resulting star formation rates
(see e.g. Figure \ref{fig:sfr_mass}).  Thus, we can state with
confidence that galaxies below this mass range grow almost entirely by
star formation, at least since $z<1$ where we focus our analysis.
This result can be understood as follows.  Halos grow via the
accretion of smaller halos.  By inspection of the lower panel of
Figure \ref{fig:theplot}, it is clear that for halos with mass
$\lesssim10^{11.5}\Msun$, corresponding to stellar masses
$\lesssim10^{10}\Msun$, the fraction of available baryons that have
been converted into stars drops precipitously.  In other words, for
these low mass halos, the even smaller mass halos that are
contributing to halo growth are almost entirely devoid of stars.
Furthermore, low mass halos have largely completed their growth by
$z\sim1$, as discussed in $\S$\ref{sec:mah}, and thus any resulting
stellar growth since $z\sim1$ must come from within the halo, i.e. via
star formation.  These points were also discussed in \citet{Purcell07}
and are qualitatively consistent with current semi-analytic models
\citep{Guo08}.  They robustly follow from the integrated star
formation efficiencies shown in Figure \ref{fig:theplot}.

At larger stellar masses the two treatments yield different
predictions for the star formation rates of galaxies.  The results
presented in $\S$\ref{sec:sfrmass} show that the data on the
SFR$-M_{\rm star}$ relation match the approach that attributes all
stellar growth to star formation, at least for stellar masses
$\lesssim10^{11}\Msun$ and $z<1$, where we focus our analysis.
Consideration of the mass range $10^{11}\lesssim M_{\rm
  star}\lesssim10^{11.5}\Msun$, where our results are somewhat less
certain (for the reasons discussed in $\S$\ref{sec:sfhg}), provides
further evidence in favor of this `no-merger' scenario, i.e., that
incoming halo mergers do not contribute substantially to the growth of
the central galaxy.  It thus appears that over this entire stellar
mass and redshift range, stellar mass growth in galaxies is dominated
by star formation.  These conclusions are largely consistent with
results from cosmological hydrodynamic simulations and may help
explain the dominance of disk galaxies at these stellar masses, if
disks are a signpost of a relatively quiescent history
\citep{Maller06}.

At first glance this may seem surprising because at these higher
masses one expects accretion of halos massive enough to host large
galaxies.  The accretion of such objects is, as mentioned above, a
generic prediction of $N$-body simulations coupled to our connection
between galaxies and halos.  Of course, the accretion of stellar
material onto the halo need not necessarily lead to growth of the
galaxy residing at the center of the halo because the accreted
material may either remain in orbit within the halo or may be tidally
disrupted before it can spiral into the center.  In the latter case,
the material will contribute to the observed diffuse intracluster
light that is ubiquitous in large dark matter halos \citep{Gonzalez05,
  Zibetti05}.  Indeed, our results indicate that some combination of
these two scenarios is precisely what is happening (see also
discussion in \citealt{Conroy07b}).  Evidence for the former scenario,
whereby accreted material remains as bound satellites, is corroborated
by the observed increase since $z\sim1$ in the fraction of galaxies at
a given halo mass that are satellites \citep{Zheng07}.

In sum, our results suggest that stellar growth since $z\sim1$ is
dominated by star formation, as opposed to mergers, for stellar masses
$\lesssim10^{11}\Msun$.  This conclusion echoes the conclusions of
\citet{Bell07} who used the observed, redshift-dependent, star
formation rate --- stellar mass relations to `predict' the evolution
of the stellar mass function since $z=1$.  These authors then compared
this predicted evolution to the actual evolution of the mass function
in order to conclude that mergers had a minor effect on the growth of
intermediate and low mass galaxies.

\section{Summary}
\label{sec:disc}

This paper presents a model for the evolution of galaxies that is
based on the observationally-motivated assumption of a tight
correlation between galaxy stellar mass and dark matter halo mass.
This assumption is used to populate halos with galaxies from $z=2$ to
$z=0$ using theoretical halo mass functions and
observationally-constrained galaxy stellar mass functions.  Halos (and
the galaxies within them) are evolved forward in time using estimates
for halo growth calibrated against $N$-body simulations.  This then
provides the average stellar mass growth of galaxies as a function of
$z=0$ stellar and halo mass.  At $M_{\rm star}\lesssim10^{10}\Msun$
the model robustly predicts that the vast majority of stellar growth
is due to in situ star formation since small halos do not accrete
significant amounts of stellar material.  At higher masses, where the
{\em halo} merger rate is higher, mergers and accretion could in
principle contribute to stellar growth.  However, the model agrees
with an array of data when all stellar growth at these higher masses
is attributed to star formation (rather than a substantial fraction
being due to mergers) for galaxies with $M_{\rm
  star}\lesssim10^{11}\Msun$ at $z<1$.  We do not use our model to
address the growth history of more massive galaxies because various
aspects of the model become uncertain in this regime.

With the assumption of a one-to-one correlation between stellar and
halo mass, the only freedom within our framework is the particular
form adopted for the redshift-dependent stellar mass function.  We
have adopted a form that provides the best fit to a variety of data
including the observed stellar mass function at $0<z<1$, the cosmic
SFR and stellar mass density, and the SFR$-M_{\rm star}$ relation over
the range $0<z<1$.  This model can thus be thought of, in part, as a
self-consistent synthesis of the available data relating galaxy SFRs
and stellar masses across time; it allows us to connect galaxy
populations at a given epoch with those at another epoch.  The model
also effectively connects the observations to the underlying dark
matter structure, thereby providing a bridge between observational
results and theoretical work aimed at understanding such observations.

The principle new result that can be obtained from this framework is a
directly-constrained form of the star formation rate in galaxies as a
function of halo mass.  Our approach provides a direct link between
observations and these models in the sense that any model which
reproduces this constrained relation for SFR$(M_{\rm vir},z)$ will
automatically match the wide variety of observational results
discussed herein, over the last half of the Universe's age.  This
result can thus help to distinguish between the processes responsible
for triggering and halting star formation, and can be directly
employed in constraining models and simulations of the physics of
galaxy formation.

The success of this simple model at describing an array of data over
the stellar mass range $10^9\lesssim M_{\rm star}\lesssim 10^{11}
\Msun$ and redshift range $0<z<1$ indicates that the relation between
galaxies and halos is surprisingly simple, smooth, and monotonic over
these ranges.  The most significant short-coming of this model is its
inability (in its present form) to predict distributions of
properties, rather than averages, as a function of stellar and halo
mass.  Such information is clearly needed to understand the color
bi-modality seen in the color-magnitude diagram, as well as the
detailed properties of satellite galaxies, and we will address this in
future work.

This model relies on observational inputs that are rather uncertain,
such as the evolution of the stellar mass function and the IMF, and
the quantitative predictions of this model are thus necessarily
uncertain.  Despite these unavoidable uncertainties, the general
trends predicted by this model, such as the dependence of the SFR of
galaxies on galaxy and halo mass, are robust and highlight the
underlying connections both between the panoply of observations at
high and low redshift, and between the observations as a whole and the
underlying dark matter distribution.

\acknowledgments 

We thank Andrew Hopkins, Kai Noeske, Ben Panter, Pablo
P\'erez-Gonz\'alez, Samir Salim, and Stephen Wilkins for providing
their data in electronic format and substantial help in its
interpretation, Kyle Stewart for providing his simulation results, and
Jeremy Tinker for generously providing his mass function and cosmology
code.  We thank Marcelo Alvarez, Peter Behroozi, Niv Drory, Sandy
Faber, Andrew Hopkins, Andrey Kravtsov, Kai Noeske, and Aristotle
Socrates for helpful conversations, and Brian Gerke, Ari Maller, Samir
Salim, and David Schiminovich for helpful comments on an earlier
draft.  RHW thanks the San Francisco skyline for inspiration; CC
thanks Princeton for being monotonic.  RHW was supported in part by
the U.S. Department of Energy under contract number DE-AC02-76SF00515
and by a Terman Fellowship at Stanford University.  We thank the Aspen
Center for Physics (partially funded by NSF-0602228) for hosting us
while much of this work was completed.  Last, but certainly not least,
we thank the referee, Eric Bell, for a careful and constructive
referee's report.


\end{document}